\documentclass[11pt,a4paper]{article}
\pdfoutput=1

\usepackage{jheppub}

\usepackage{amsmath, amssymb, amsthm} 
\usepackage{mathtools}
\usepackage{thmtools}
\usepackage{bm}
\usepackage{dsfont}
\usepackage{braket}
\usepackage{graphicx}

\numberwithin{equation}{section}

\usepackage{enumitem}
\setlist[itemize]{noitemsep}
\setlist[description]{noitemsep}

\usepackage[dvipsnames]{xcolor}
\usepackage{tikz}
\usepackage{pgfplots}
\usetikzlibrary{intersections,backgrounds}
\usepgfplotslibrary{fillbetween}
\pgfplotsset{compat = newest}

\usepackage[numbers, sort&compress]{natbib}
\usepackage{hyperref}
\hypersetup{colorlinks=true, linktoc=page, linkcolor=purple, citecolor=blue}

\newcommand{\floor}[1]{\cramped{\left\lfloor #1 \right\rfloor}}

\newcommand{\negphantom}[1]{
    \ifmmode\settowidth{\dimen0}{$#1$}
    \else\settowidth{\dimen0}{#1}
    \fi
    \hspace*{-\dimen0}}
    
\makeatletter
\newcommand{\mask}[2]{{\mathpalette\mask@{{#1}{#2}}}}
\newcommand{\mask@}[2]{\mask@@{#1}#2}
\newcommand{\mask@@}[3]{%
  \settowidth{\dimen@}{$\m@th#1#2$}%
  \makebox[\dimen@]{$\m@th#1#3$}%
}
\makeatother
    
\newcommand{\sm}{\smallskip}

\newcommand{\fsl}[1]{\ensuremath{\mathrlap{\!\not{\phantom{#1}}}#1}}

\usepackage{tikz}
\usetikzlibrary{decorations.pathmorphing,decorations.shapes}
\usetikzlibrary{decorations.pathreplacing,decorations.markings}
\usetikzlibrary{backgrounds}
\usetikzlibrary{positioning}
\usetikzlibrary{arrows}
\usetikzlibrary{shapes,shapes.geometric,shapes.misc}

\tikzstyle{tikzfig}=[baseline=-0.25em,scale=0.5]
\pgfkeys{/tikz/tikzit fill/.initial=0}
\pgfkeys{/tikz/tikzit draw/.initial=0}
\pgfkeys{/tikz/tikzit shape/.initial=0}
\pgfkeys{/tikz/tikzit category/.initial=0}

\pgfdeclarelayer{edgelayer}
\pgfdeclarelayer{nodelayer}
\pgfsetlayers{background,edgelayer,nodelayer,main}

\tikzstyle{none}=[inner sep=0mm]

\newcommand{\tikzfig}[1]{%
{\tikzstyle{every picture}=[tikzfig]
\IfFileExists{#1.tikz}
  {\input{#1.tikz}}
  {%
    \IfFileExists{./figures/#1.tikz}
      {\input{./figures/#1.tikz}}
      {\tikz[baseline=-0.5em]{\node[draw=red,font=\color{red},fill=red!10!white] {\textit{#1}};}}%
  }}%
}

\tikzset{->-/.style={decoration={
  markings,
  mark=at position #1 with {\arrow{>}}},postaction={decorate}}}

\tikzset{-<-/.style={decoration={
  markings,
  mark=at position #1 with {\arrow{<}}},postaction={decorate}}}

\tikzstyle{every loop}=[]
\begin{document}

\title{Consistent actions for massive particles interacting with electromagnetism and gravity}

\date{\today}

\author[a]{Lukas W. Lindwasser}

\affiliation[a]{
    Mani L. Bhaumik Institute for Theoretical Physics\\
    Department of Physics and Astronomy\\
    University of California, Los Angeles, CA 90095, USA}

\emailAdd{lukaslindwasser@physics.ucla.edu}


\abstract{Consistent interactions with electromagnetism and gravity for mass $m$ particles of any spin are obtained. This is done by finding interactions which preserve the covariantized massive gauge symmetry present in recently constructed massive particle actions. This gauge principle is sufficient for finding consistent completions of minimal as well as non-minimal couplings of any type. For spins $s\geq 3/2$, consistency requires infinitely many interaction terms in the action, including arbitrarily high order derivatives of electromagnetic and gravitational curvatures, with correspondingly high powers of $1/m$. These interactions may be formally resummed and expressed in terms of non-local operators. Finally, although the interactions appear non-local, evidence is presented for the existence of a field redefinition which makes the interacting action local. This work provides the first explicit realization of an exactly gauge invariant formulation of massive particles interacting with electromagnetism and gravity.}


\maketitle

 \newpage

\section{Introduction}
\label{sec:intro}
There exists in nature composite massive particles with spins higher than 1, the highest spin particle observed at the moment being the $\Delta(2950)$ baryon with spin 15/2 \cite{ParticleDataGroup:2022pth}. The basic problem of modelling such particles interacting with the most relevant forces they experience during their lifetime, electromagnetism and gravity, has a long history, first considered by Fierz and Pauli in 1939 \cite{Fierz:1939ix}. In that work, they noted that introducing interactions generically does not preserve the massive higher spin particle's degrees of freedom, counted by considering the number of constraints offered from the equations of motion. 

\sm

Another challenge to the consistent modelling of interacting massive higher spin particles is that interactions violate causality. For instance, it was realized by Velo and Zwanziger in 1969 \cite{PhysRev.186.1337} that minimally coupling spin 3/2 particles to an electromagnetic background allows for superluminal propagation. Subsequently, arguments \cite{Camanho:2014apa,Afkhami-Jeddi:2018apj} were made that any theory with a finite collection of interacting massive particles of spin higher than 2 cannot avoid causality violations at sufficiently high energies. For spin 3/2 particles, a solution to this and the degree of freedom problem was found by realizing the particle within the framework of supergravity \cite{Deser:1977uq}. For higher spins, the only known solution to both is to realize them within string theory \cite{Porrati:2010hm}, which has an infinite collection of higher spin particle excitations. 

\sm

Barring these special solutions, these results are usually taken to mean that massive higher spin particles cannot be elementary \cite{Arkani-Hamed:2017jhn,Afkhami-Jeddi:2018apj}. This is related to the fact that tree level scattering amplitudes involving massive higher spin particles have bad high energy behavior which violate expected bounds. It is known that there are unique non-minimal couplings with electromagnetism and gravity that improves the bad high energy behavior in some, but importantly not all, scattering processes \cite{Ferrara:1992yc,Cucchieri:1994tx}. Furthermore, loop diagrams involving particles with spin 3/2 and higher result in nonrenormalizeable divergences. At energies where these bounds saturate, the particle description must be replaced with some yet more fundamental description. Indeed, all massive particles/resonances with spin 3/2 and higher we have observed are understood to be composites of low spin particles. For these reasons, modelling interacting massive higher spin particles directly is firmly within the regime of effective field theory.

\sm

At the moment, there is an active program of research modelling the inspiral of two spinning black holes or neutron stars using effective field theory techniques, treating them as point particles with very high spin \cite{Guevara:2017csg,Chung:2018kqs, Guevara:2018wpp, Arkani-Hamed:2019ymq,Guevara:2019fsj,Bern:2020buy, Bern:2020uwk, Bern:2022kto, Chiodaroli:2021eug, FebresCordero:2022jts, Aoude:2022thd, Cangemi:2022bew}, in order to explain recently detected gravitational waves \cite{LIGOScientific:2016aoc,LIGOScientific:2017vwq,LIGOScientific:2021djp}. This has reinvigorated research in the modelling of interacting massive higher spin particles.

\sm

A low energy effective field theory with an energy cutoff $\Lambda$ starts to see the aforementioned bad high energy behavior at energies higher than the higher spin particle's mass $m$. Causality violations meanwhile are suppressed but still detectable if a superluminal signal with speed difference $\Delta v = v-c > 0$ propagates for a sufficiently long time $T$ such that $\Delta v T\gtrsim 1/\Lambda$, or again if $\Lambda$ is parametrically larger than the higher spin particle's mass. Indeed, positivity bounds suggest that the energy cutoff cannot be much larger than the mass $\Lambda\sim m$, with a gap which decreases as the spin increases \cite{Bellazzini:2019bzh,Bellazzini:2023nqj}, putting strong constraints on the range of validity of any effective field theory of massive higher spin particles in the limit $s\to\infty$. Still, one expects to be able to model any classically spinning object as a highly spinning point particle at sufficiently long length scales, and so there remains hope that an effective field theory approach for such scenarios can be made sensible.

\sm

Because we are strictly within the regime of effective field theory, the action for a massive higher spin particle should at best be understood perturbatively as a generating functional of the $S$-matrix, and other observables. As long as the resulting $S$-matrix is well defined in the expected range of validity of the theory, the asymptotic states are chosen from the free part of the action, after any spontaneous symmetry breaking or Higgsing is taken into account. Any further inconsistencies should inform the actual range of validity of the theory. 

\sm

Using the free Singh-Hagen actions \cite{Singh:1974qz,Singh:1974rc} for a mass $m$ spin $s$ particle as a starting point for this effective field theory, and adding interactions, one encounters technical challenges when computing scattering amplitudes. These actions use a traceless symmetric rank $n$ tensor $\phi_{\mu_1\cdots\mu_n}$ for integer spin $s=n$, and a $\gamma$ traceless and symmetric rank $n$ tensor $\psi_{\mu_1\cdots\mu_n}$ for half integer spin $s=n+1/2$, which are accompanied by several auxiliary fields $\phi_{\mu_1\cdots\mu_k}$, $\psi_{\mu_1\cdots\mu_k}$ of lower rank which intricately couple to impose the mass constraints $(\partial^2-m^2)\phi_{\mu_1\cdots\mu_n}=0$, $(\fsl{\partial}+m)\psi_{\mu_1\cdots\mu_n}=0$, and transversality constraints $\partial^{\mu_1}\phi_{\mu_1\cdots\mu_n}=0$, $\partial^{\mu_1}\psi_{\mu_1\cdots\mu_n}=0$, all while making the auxiliary fields vanish. The propagator for the physical spin $s$ field is known to grow like $\sim p^{-2+2s}$ in momentum space \cite{Weinberg:1964cn, Singh:1981aw}. Because the auxiliary fields vanish when the equations of motion are imposed, their propagators do not in general vanish but are instead at most a polynomial in momentum space, and must be included in your Feynman rules. These polynomials are complicated and are in fact not known for all spins, making a large spin limit for the purposes of modelling classically spinning objects not at the moment possible without guessing a pattern with low spin examples and extrapolating. It is important to develop tools which facilitate a large spin limit without extrapolation.

\sm

Instead of using the Singh-Hagen actions, one can use a different formulation of massive higher spin particles for which the propagators are known for all spins. This can be achieved in the various gauge invariant formulations of massive higher spin particles known in the literature \cite{Klishevich:1998ng,Klishevich:1998yt,Zinoviev:2001dt,Metsaev:2006zy,Asano:2019smc,Lindwasser:2023zwo}, which incorporate more auxiliary fields and a compensating gauge symmetry, which allows the freedom to pick a gauge in which the propagators are easily computable as a function of the spin. In this paper, we will focus on the gauge invariant formulation presented in \cite{Lindwasser:2023zwo}, for which there is a known gauge where the propagators involving the physical spin $s$ field as well as all auxiliary fields are known explicitly as a function of $s$ and the spacetime dimension $d$, which behave like $\sim p^{-2}$ and $\sim p^{-1}$ in momentum space for all integer spins and half integer spins, respectively.

\sm

With the explicit propagators known in a particular gauge, the challenge then becomes writing down interactions which are gauge invariant, so that the choice of gauge does not effect observables. An additional challenge is to find gauge invariant interactions which may be written in a uniform manner for any spin, so that an explicit large spin limit of observables is foreseeable. There is a reliable perturbative procedure for introducing interactions consistent with the gauge symmetry, while also possibly deforming the gauge transformations, described for instance in \cite{Berends:1984rq} and references therein. This approach is known as the Noether procedure, and provides a way to construct for instance Yang-Mills theory, general relativity, and supergravity, starting from a collection of free fields and free gauge symmetries. Any set of interactions modulo those removable via field redefinition which results from an exact solution to this perturbative procedure is referred to as ``consistent" \cite{Barnich:1993vg}. In this paper, we will focus on constructing consistent interactions between massive particles of any spin, and electromagnetism and gravity. 

\sm

Progress in this direction has been made in the past, including work at the level of equations of motion \cite{Porrati:2010hm,Cortese:2013lda,Rahman:2016tqc,Rahman:2020qal}, at the Lagrangian level for constrained external backgrounds \cite{Argyres:1989cu,Porrati:2009bs,Buchbinder:2000fy,Zinoviev:2006im,Zinoviev:2008ck,Zinoviev:2009hu,Buchbinder:2012iz}, and for weak interactions at cubic order  \cite{Metsaev:2012uy,Metsaev:2022yvb}. An important challenge is to find consistent actions of massive higher spin fields interacting in an arbitrary electromagnetic and gravitational field, so that they too can be treated as dynamical\footnote{A notable work is \cite{Ochirov:2022nqz}, which uses a chiral representation $(s,0)$ of the Lorentz group rather than $(s/2,s/2)$. Further care must be taken if one wants a parity invariant theory.}. The general expectation is that an exactly gauge invariant theory of massive higher spins with electromagnetism and gravity will require infinitely many interactions, including all orders of derivatives of curvatures, and correspondingly all orders in $1/m$. In this paper, a prescription for writing down all such interactions is found for any spin.

\sm

The actions $S_n$ and $S_{n+1/2}$ constructed in \cite{Lindwasser:2023zwo} are built from gauge invariant field strengths $\mathcal{F}_{n-i}$, $i=0,1,2,3$ for integer spins and $\mathcal{S}_{n-i}$, $i = 0,1,2$ for half integers, respectively, which satisfy massive Bianchi identities\footnote{The name ``massive Bianchi identity" comes from the fact that they descend from massless higher spin identities, explained in \cite{Lindwasser:2023zwo}, which are analogous to the linearized second contracted Bianchi identity in Riemannian geometry when the spin $s=n=2$.} (\ref{eq:intBianch1}), (\ref{eq:intBianch2}), (\ref{eq:halfBianch}), enabling a simple prescription for finding exactly gauge invariant interactions. While the general procedure for adding consistent interactions \cite{Berends:1984rq} entails also deforming the free theory's gauge transformations, possibly including terms proportional to arbitrary powers of electromagnetic and gravitational curvatures $F_{\mu\nu}$ and $R_{\omega\sigma\mu\nu}$, we find it sufficient to minimally deform the gauge transformations by simply replacing spacetime derivatives present in them with covariant derivatives, so that it is at least consistent with the $U(1)$ electromagnetic gauge and general coordinate invariance. Given this gauge principle, the gauge variation of the actions $\delta S_n$ and $\delta S_{n + 1/2}$ depends only on the aforementioned properties of the field strengths. 

\sm

Minimally coupling the massive higher spin fields $\Phi_{n-i},\,\Psi_{n-i}$ to electromagnetism and gravity modifies the field strengths so that they are no longer gauge invariant~$\delta\mathcal{F}_{n-i}$, $\delta\mathcal{S}_{n-i}\neq 0$, and they no longer satisfy massive Bianchi identities. This in turn breaks the gauge invariance of the actions $\delta S_n,\,\delta S_{n+1/2}\neq 0$. Because of the way in which the gauge symmetry is broken, the gauge symmetry can be restored by further modifying the field strengths $\mathcal{F}_{n-i}\to \mathcal{F}_{n-i} + \Delta\mathcal{F}_{n-i}$ and $\mathcal{S}_{n-i}\to \mathcal{S}_{n-i} + \Delta\mathcal{S}_{n-i}$ so that they again are gauge invariant $\delta(\mathcal{F}_{n-i} + \Delta\mathcal{F}_{n-i}),\,\delta(\mathcal{S}_{n-i} + \Delta\mathcal{S}_{n-i}) = 0$, and satisfy massive Bianchi identities. Important for the ability to perform a large spin limit, this specific setup for adding consistent interactions will result in expressions valid for all spins. $\Delta\mathcal{F}_{n-i}$ and $\Delta\mathcal{S}_{n-i}$ remain linear in $\Phi_{n-i}$ and $\Psi_{n-i}$, respectively, but include powers of $F_{\mu\nu}$ and $R_{\omega\sigma\mu\nu}$ and all powers of their covariant derivatives. More general interactions, corresponding for instance to multipole and tidal couplings classified in \cite{Bern:2020buy,Bern:2020uwk}, can be made gauge invariant in the same way. The modifications $\Delta\mathcal{F}_{n-i}$ and $\Delta\mathcal{S}_{n-i}$ are generically non-local in nature, with non-local length scale $L\sim 1/m$. Importantly, we will find that the interactions presented in this paper cannot be removed altogether via some field redefinition, and we present evidence for the existence of a field redefinition which removes the non-locality from the action. 

\sm

Despite such a field redefinition, it appears that the energy cutoff remains at $\Lambda\sim m$.  An optimistic bound on the energy cutoff of a mass $m$ spin $s$ particle interacting with gravity is conjectured to be $\Lambda_s\sim(m^{2s-2}M_{pl})^{\frac{1}{2s-1}}$, where $M_{pl}$ is the Planck mass \cite{Rahman:2009zz}. We arrived at a cutoff $\Lambda = \lim_{s\to\infty}\Lambda_s$ because of our insistence in making the interactions gauge invariant in a manner independent of spin to facilitate a large spin limit.
\subsection{Conventions}
Throughout this paper, we use the metric signature $\eta = \text{diag}(-1,+1,\cdots,+1)$. In terms of the $U(1)$ gauge field $A_{\mu}$ and Levi-Civita connection $\Gamma^{\lambda}_{\;\;\mu\nu}$, our convention for electromagnetic and gravitational curvatures are
\begin{align}
    &F_{\mu\nu} = \partial_{\mu}A_{\nu} - \partial_{\nu}A_{\mu} \\
    &R^{\lambda}_{\;\;\sigma\mu\nu} = \partial_{\mu}\Gamma^{\lambda}_{\;\;\nu\sigma} - \partial_{\nu}\Gamma^{\lambda}_{\;\;\mu\sigma} + \Gamma^{\lambda}_{\;\;\mu\kappa}\Gamma^{\kappa}_{\;\;\nu\sigma} - \Gamma^{\lambda}_{\;\;\nu\kappa}\Gamma^{\kappa}_{\;\;\mu\sigma}
\end{align}
Finally, we will on occasion write $\gamma_{\mu\nu}\equiv\frac{1}{2}[\gamma_{\mu},\gamma_{\nu}]$ for shorthand, where $\gamma_{\mu}$ are Dirac matrices satisfying a Clifford algebra for either the flat $\eta_{\mu\nu}$ or curved $g_{\mu\nu}$ metric depending on the context.
\subsection{Outline}
In \autoref{sec:free}, we review the hyperfield formalism that will help simplify the subsequent analysis, and the free actions for massive integer and half integer particles constructed in \cite{Lindwasser:2023zwo}. In \autoref{sec:inter}, we present exactly gauge invariant interactions between massive integer spin particles and electromagnetism and gravity in \autoref{sec:integer}, and then repeat the analysis for half integer spin particles in \autoref{sec:halfinteger}. In \autoref{sec:fieldredef}, we discuss an interpretation of the resulting non-local interactions, providing evidence that the actions are equivalent to local actions via some field redefinition. Importantly, we demonstrate that the interactions constructed in this paper cannot be removed altogether via some field redefinition, and hence have a physical effect on observables. Finally in \autoref{sec:disc}, we discuss our results and conclude with some directions for future research. 

\section{Free theory}
\label{sec:free}
In this section, we briefly review the formulation of free massive spinning particles \cite{Lindwasser:2023zwo} that will be used as our starting point for incorporating interactions. 

\sm

Particles with spin are covariantly described by totally symmetric tensors $\phi_{\mu_1\cdots\mu_n}(x)$ for spin $n$, or spinor tensors $\psi_{\mu_1\cdots\mu_n}(x)$ for spin $n+1/2$. The formulation of these particles dramatically simplifies after the introduction of an auxiliary vector coordinate $s^{\mu}$, and instead writing the theory in terms of a ``hyperfield" $\Phi_n(X,s) = \frac{1}{n!}i^{-n/2}\phi_{\mu_1\cdots\mu_n}(X)s^{\mu_1}\cdots s^{\mu_n}$, where the factor $i^{-n/2}$ is added for later convenience. All of the basic operations needed for constructing the covariant actions for massive spinning particles can be performed at the level of the hyperfield. For instance, one often has to take traces of the fields $\phi^{\lambda}_{\;\;\lambda\mu_3\cdots\mu_n}(X)$. This can be achieved at the level of $\Phi_n(X,s)$ by taking the Laplacian with respect to $s^{\mu}$
\begin{align}
    \label{eq:trace}
   & i\,\partial_s^2\Phi_n(X,s) = \frac{1}{(n-2)!}i^{-(n-2)/2}\phi^{\lambda}_{\;\;\lambda\mu_1\cdots\mu_{n-2}}(X)s^{\mu_1}\cdots s^{\mu_{n-2}} 
   \intertext{The divergence $\partial^{\lambda}\phi_{\lambda\mu_2\cdots\mu_n}(X)$ may also be written in terms of $\Phi_n(X,s)$}
   \label{eq:diverge}
    & i^{1/2}\,\partial_s\cdot\partial_X\Phi_n(X,s) = \frac{1}{(n-1)!}i^{-(n-1)/2}\partial^{\lambda}\phi_{\lambda\mu_1\cdots\mu_{n-1}}(X)s^{\mu_1}\cdots s^{\mu_{n-1}}
   \intertext{The symmetric derivative $\partial_{(\mu_1}\phi_{\mu_2\cdots\mu_{n+1})}(X)$ is written in terms of $\Phi_n(X,s)$ via}
   \label{eq:symmderiv}
    & i^{-1/2}s\cdot\partial_X\Phi_n(X,s) = \frac{1}{n!}i^{-(n+1)/2}\partial_{(\mu_1}\phi_{\mu_2\cdots\mu_{n+1})}(X)s^{\mu_1}\cdots s^{\mu_{n+1}}
   \intertext{One can also contract indices between two equal rank $n$ hyperfields $A_n(X,s)$ and $B_n(X,s)$}
   \label{eq:contract}
    & \int \frac{d^dsd^ds'}{(2\pi)^d}e^{is\cdot s'}A_n(X,s)B_n(X,s') = \frac{1}{n!}a_{\mu_1\cdots\mu_n}b^{\mu_1\cdots\mu_n}
   \intertext{These four operations are sufficient for the formulation of integer spin fields. For half integer spin fields, contraction with gamma matrices $\gamma^{\mu}$ will also be necessary. For a Dirac hyperfield $\Psi_n(X,s)$, we have}
   \label{eq:gammatrace}
    & i^{1/2}\fsl{\partial}_s\Psi_n(X,s) = \frac{1}{(n-1)!}i^{-(n-1)/2}\gamma^{\lambda}\psi_{\lambda\mu_1\cdots\mu_n}(X) s^{\mu_1}\cdots s^{\mu_n}
\end{align}

\sm

The formula (\ref{eq:contract}) is particularly interesting, as it suggests introducing a (pseudo) inner product on the space of hyperfields
\begin{align}
\label{eq:hyperprod}
    &(A_n,B_n) = \int d^dX\frac{d^dsd^ds'}{(2\pi)^d}e^{is\cdot s'}\tilde{A}_n(X,s)B_n(X,s') = \int d^dX \frac{1}{n!}a^*_{\mu_1\cdots\mu_n}b^{\mu_1\cdots\mu_n} \\
    &\text{where } \tilde{A}_n(X,s) = \frac{1}{n!}i^{-n/2}a^*_{\mu_1\cdots\mu_n}(X)s^{\mu_1}\cdots s^{\mu_n}
\end{align}
and we may consider the (pseudo) Hilbert space of hyperfields $\Phi(X,s)$ with finite (pseudo) norm $(\Phi,\Phi)<\infty$\footnote{$(A,B)$ is only positive definite in Euclidean signature.} \cite{Bargmann:1977gy}. Note that $(i^{-1/2}s^{\mu}A,B)=(A,i^{1/2}\partial_s^{\mu}B)$, and so for instance, the divergence and symmetric derivative are anti-Hermitian adjoints of each other~$(i^{1/2}\partial_s\cdot\partial_X)^\dagger = -i^{-1/2}s\cdot\partial_X$ in this space. 

\subsection{Integer spins}
As detailed in \cite{Lindwasser:2023zwo}, the covariant formalism of a complex massive spin $n$ particle we work with not only uses a rank $n$ hyperfield $\Phi_n(X,s)$, but also three auxiliary hyperfields $\Phi_{n-1}(X,s)$, $\Phi_{n-2}(X,s)$, and $\Phi_{n-3}(X,s)$. These auxiliary hyperfields couple to $\Phi_n(X,s)$ in such a way so that after imposing the equations of motion, the auxiliary hyperfields may be set to zero, and $\Phi_n(X,s)$ is a Fierz-Pauli system \cite{Fierz:1939ix} in hyperspace
\begin{align}
\label{eq:hyperFP}
    (\partial_X^2-m^2)\Phi_n &= 0 \nonumber\\
    \partial_s\cdot\partial_X\Phi_n &=0 \nonumber\\
    \partial_s^2\Phi_n &=0 
\end{align}
$\Phi_n(X,s)$ therefore has the correct degrees of freedom for the description of a massive spin $n$ particle, with mass $m$. This special coupling is facilitated by demanding that the theory be invariant under the gauge transformations
\begin{align}
\label{eq:Stueck}
    &\delta\Phi_n = i^{-1/2} s\cdot\partial_X \epsilon_{n-1} && \delta\Phi_{n-1} = i^{-1/2} s\cdot \partial_X \epsilon_{n-2} + im\,\epsilon_{n-1} \nonumber\\
    &\delta\Phi_{n-2} = -i^{1/2} s\cdot\partial_X\partial_s^2\epsilon_{n-1} +2im\,\epsilon_{n-2} && \delta\Phi_{n-3} = -i^{1/2} s\cdot\partial_X\partial_s^2\epsilon_{n-2} +3m\partial_s^2\epsilon_{n-1}
\end{align}
where the gauge parameters $\epsilon_{n-1}(X,s)$ and $\epsilon_{n-2}(X,s)$ are arbitrary rank $n-1$ and rank $n-2$ hyperfields, respectively.

\sm

This theory includes gauge invariant field strengths $\mathcal{F}_n$, $\mathcal{F}_{n-1}$, $\mathcal{F}_{n-2}$, and $\mathcal{F}_{n-3}$, which are linear in the $\Phi_{n-i}$'s and quadratic in derivatives
\begin{align}
    \mathcal{F}_n =& \big(\partial_X^2-m^2-s\cdot\partial_X\partial_s\cdot\partial_X+\frac{1}{2}(s\cdot\partial_X)^2\partial_s^2\big)\Phi_n-\frac{i}{2}(s\cdot\partial_X)^2\Phi_{n-2} - i^{1/2}m \,s\cdot\partial_X\Phi_{n-1} \\
    \mathcal{F}_{n-1} =& \big(\partial_X^2-s\cdot\partial_X\partial_s\cdot\partial_X+\frac{1}{2}(s\cdot\partial_X)^2\partial_s^2\big)\Phi_{n-1} - \frac{i}{2}(s\cdot\partial_X)^2\Phi_{n-3}\nonumber \\
    & + i^{-1/2}m\partial_s\cdot\partial_X\Phi_n - i^{-1/2} m\, s\cdot\partial_X\partial_s^2\Phi_n \\
    \mathcal{F}_{n-2} =& \big(\partial_X^2-s\cdot\partial_X\partial_s\cdot\partial_X-\frac{1}{2}(s\cdot\partial_X)^2\partial_s^2\big)\Phi_{n-2}- \frac{i}{2} (s\cdot\partial_X)^2\partial_s^4\Phi_n - im^2\partial_s^2\Phi_n \nonumber\\
    & + 2i^{-1/2}m\partial_s\cdot\partial_X\Phi_{n-1} - 2i^{-1/2} m\, s\cdot\partial_X\partial_s^2\Phi_{n-1} + i^{1/2}m\, s\cdot\partial_X\Phi_{n-3} \\
    \mathcal{F}_{n-3}=&\big(\partial_X^2-m^2-s\cdot\partial_X\partial_s\cdot\partial_X-\frac{1}{2}(s\cdot\partial_X)^2\partial_s^2\big)\Phi_{n-3} -\frac{i}{2}(s\cdot\partial_X)^2\partial_s^4\Phi_{n-1} - 3im^2\partial_s^2\Phi_{n-1} \nonumber \\
    & +2i^{1/2}m\, s\cdot\partial_X\partial_s^4\Phi_n +3i^{-1/2}m\partial_s\cdot\partial_X\Phi_{n-2} + i^{-1/2} m\, s\cdot\partial_X\partial_s^2\Phi_{n-2}
\end{align}
The equations of motion are equivalent to setting all four field strengths to zero $\mathcal{F}_{n-i}=0$. In terms of these ingredients, the gauge invariant free action for a massive spin $n$ particle may be written as
\begin{align}
\label{eq:mhyperaction}
     S_n = \frac{1}{2}&n!\int d^dX\frac{d^dsd^ds'}{(2\pi)^d}e^{is\cdot s'}\times \nonumber\\
     \Bigg\{ &\sum_{k = 0}^{\floor{n/2}} \frac{(-1)^k}{(2k)!}\Bigg(&&\hspace{-2.cm}\Big(1-\frac{3k}{2}\Big)\partial_{s}^{2k}\tilde{\Phi}_n\partial_{s'}^{2k}\mathcal{F}_n+i\frac{k}{2}\partial_{s}^{2k}\tilde{\Phi}_n\partial_{s'}^{2(k-1)}\mathcal{F}_{n-2} \nonumber\\
    & && \hspace{-1.5cm}+i\frac{k}{2}\partial_{s}^{2(k-1)}\tilde{\Phi}_{n-2}\partial_{s'}^{2k}\mathcal{F}_n -\frac{k}{2}\partial_{s}^{2(k-1)}\tilde{\Phi}_{n-2}\partial_{s'}^{2(k-1)}\mathcal{F}_{n-2}\Bigg) \nonumber\\
    & \hspace{-.5 cm}+ \sum_{k = 0}^{\floor{(n-1)/2}} \frac{(-1)^k}{(2k+1)!}\Bigg(&&\hspace{-1cm}\Big(1-\frac{5k}{2}\Big)\partial_{s}^{2k}\tilde{\Phi}_{n-1}\partial_{s'}^{2k}\mathcal{F}_{n-1}+i\frac{3k}{2}\partial_{s}^{2k}\tilde{\Phi}_{n-1}\partial_{s'}^{2(k-1)}\mathcal{F}_{n-3} \nonumber \\
    & &&\hspace{-1.15cm}+i\frac{3k}{2}\partial_{s}^{2(k-1)}\tilde{\Phi}_{n-3}\partial_{s'}^{2k}\mathcal{F}_{n-1} +\frac{k}{2}\partial_{s}^{2(k-1)}\tilde{\Phi}_{n-3}\partial_{s'}^{2(k-1)}\mathcal{F}_{n-3}\Bigg)\Bigg\} \nonumber \\
    & &&\hspace{7cm} + \text{c.c.}
\end{align}
That this action is gauge invariant is guaranteed by the fact that the field strengths are themselves gauge invariant $\delta\mathcal{F}_{n-i}=0$, and that they satisfy what we call massive Bianchi identities
\begin{align}
\label{eq:intBianch1}
    &\partial_s\cdot\partial_X\mathcal{F}_n-\frac{1}{2}s\cdot\partial_X\partial_s^2\mathcal{F}_n + \frac{i}{2}s\cdot\partial_X\mathcal{F}_{n-2} + i^{1/2}m\mathcal{F}_{n-1} = 0   \\
\label{eq:intBianch2}
    &\partial_s\cdot\partial_X\mathcal{F}_{n-1} - \frac{1}{2}s\cdot\partial_X\partial_s^2\mathcal{F}_{n-1} + \frac{i}{2}s\cdot\partial_X\mathcal{F}_{n-3} +\frac{1}{2}i^{-1/2}m\partial_s^2\mathcal{F}_n +\frac{1}{2}i^{1/2}m\mathcal{F}_{n-2} = 0 
\end{align}
Indeed, the gauge variation of the action $\delta S_n$ is a straightforward linear combination of terms proportional to the left hand sides of (\ref{eq:intBianch1}), (\ref{eq:intBianch2}), and the $\delta\mathcal{F}_{n-i}$'s, and their complex conjugates. This fact will be very important for us when considering interactions in \autoref{sec:inter}.
\subsection{Half integer spins}
The covariant formalism for Dirac spin $n+1/2$ particles \cite{Lindwasser:2023zwo} is very analogous to the integer spin case. It requires the use of not only a rank $n$ Dirac spinor hyperfield $\Psi_n(X,s)$, but also two auxiliary Dirac spinor hyperfields $\Psi_{n-1}(X,s)$ and $\Psi_{n-2}(X,s)$. As before, these auxiliary hyperfields couple to $\Psi_n(X,s)$ in such a way so that after imposing the equations of motion, the auxiliary hyperfields may be set to zero, and $\Psi_n(X,s)$ is a Rarita-Schwinger system \cite{PhysRev.60.61} in hyperspace
\begin{align}
\label{eq:hyperRS}
    (\fsl{\partial}_X + m)\Psi_n&=0\nonumber\\
    \fsl{\partial}_s\Psi_n&=0
\end{align}
$\Psi_n(X,s)$ therefore has the correct degrees of freedom for the description of a massive spin $n + 1/2$ particle, with mass $m$. This special coupling is facilitated by demanding that the theory be invariant under the gauge transformations
\begin{align}
\label{eq:halfStueck}
&\delta\Psi_n=i^{-1/2}s\cdot\partial_X\epsilon_{n-1} \nonumber\\
&\delta\Psi_{n-1} = -i s\cdot\partial_X\fsl{\partial}_s\epsilon_{n-1}+im\epsilon_{n-1} \nonumber\\
&\delta\Psi_{n-2} = -i^{1/2}s\cdot\partial_X\partial_s^2\epsilon_{n-1}+2i^{1/2}m\fsl{\partial}_s\epsilon_{n-1}
\end{align}
where the gauge parameter $\epsilon_{n-1}(X,s)$ is an arbitrary rank $n-1$ hyperfield.

\sm

This theory includes gauge invariant field strengths $\mathcal{S}_n$, $\mathcal{S}_{n-1}$, and $\mathcal{S}_{n-2}$, which are linear in the $\Psi_{n-i}$'s and linear in derivatives 
\begin{align}
    \mathcal{S}_n &= (\fsl{\partial}_X+m-s\cdot\partial_X\fsl{\partial}_s)\Psi_n+i^{1/2}s\cdot\partial_X\Psi_{n-1} \\
    \mathcal{S}_{n-1} &= (\fsl{\partial}_X-s\cdot\partial_X\fsl{\partial}_s)\Psi_{n-1} + i^{1/2}s\cdot\partial_X\Psi_{n-2} + i^{-1/2}m\fsl{\partial}_s\Psi_n \\
    \mathcal{S}_{n-2} &= (\fsl{\partial}_X-m)\Psi_{n-2}+is\cdot\partial_X\fsl{\partial}_s^3\Psi_n + i^{-1/2}s\cdot\partial_X\partial_s^2\Psi_{n-1} + 2i^{-1/2}m\fsl{\partial}_s\Psi_{n-1}
\end{align}
The equations of motion are equivalent to setting all three field strengths to zero $\mathcal{S}_{n-i}=0$. In terms of these ingredients, the gauge invariant free action for a massive spin $n+1/2$ particle in the Dirac representation may be written as
\begin{align}
\label{eq:mhalfhyperaction}
S_{n+1/2} &=- n!\int d^dX\frac{d^dsd^ds'}{(2\pi)^d}e^{is\cdot s'}\times\nonumber \\
\Bigg\{ &\sum_{k = 0}^{\floor{n/2}} \frac{(-1)^k}{(2k)!}\Bigg(&&\hspace{-2.75cm}\Big(1-\frac{3k}{2}\Big)\partial_{s}^{2k}\overline{\Psi}_n\partial_{s'}^{2k}\mathcal{S}_n+i\frac{k}{2}\partial_{s}^{2k}\overline{\Psi}_n\partial_{s'}^{2(k-1)}\mathcal{S}_{n-2}\nonumber\\
    & && \hspace{-3cm}+i\frac{k}{2}\partial_{s}^{2(k-1)}\overline{\Psi}_{n-2}\partial_{s'}^{2k}\mathcal{S}_n +\frac{k}{2}\partial_{s}^{2(k-1)}\overline{\Psi}_{n-2}\partial_{s'}^{2(k-1)}\mathcal{S}_{n-2}\nonumber\\
    & && \hspace{-3.5cm} -i^{-1/2}k\partial_s^{2(k-1)}\overline{\Psi}_{n-1}\overleftarrow{\fsl{\partial}_s}\partial_{s'}^{2(k-1)}\mathcal{S}_{n-2} + i^{-1/2}k\partial_s^{2(k-1)}\overline{\Psi}_{n-2}\fsl{\partial}_{s'}\partial_{s'}^{2(k-1)}\mathcal{S}_{n-1} \nonumber \\
    & && \hspace{2.75cm}+ ik\partial_{s}^{2(k-1)}\overline{\Psi}_{n-1}\overleftarrow{\fsl{\partial}_s}\fsl{\partial}_{s'}\partial_{s'}^{2(k-1)}\mathcal{S}_{n-1}\Bigg)
    \nonumber\\
    & \hspace{-1cm}+ \sum_{k = 0}^{\floor{(n-1)/2}} \frac{(-1)^k}{(2k+1)!}\Bigg(&&\hspace{-2.5cm}-i\Big(\frac{1}{2}+\frac{3k}{2}\Big)\partial_{s}^{2k}\overline{\Psi}_n\overleftarrow{\fsl{\partial}_s}\fsl{\partial}_{s'}\partial_{s'}^{2k}\mathcal{S}_n-\frac{k}{2}\partial_{s}^{2k}\overline{\Psi}_n\overleftarrow{\fsl{\partial}_s}\fsl{\partial}_{s'}\partial_{s'}^{2(k-1)}\mathcal{S}_{n-2} \nonumber \\
    & &&\hspace{-2.5cm}-\frac{k}{2}\partial_{s}^{2(k-1)}\overline{\Psi}_{n-2}\overleftarrow{\fsl{\partial}_s}\fsl{\partial}_{s'}\partial_{s'}^{2k}\mathcal{S}_n + i\frac{k}{2}\partial_{s}^{2(k-1)}\overline{\Psi}_{n-2}\overleftarrow{\fsl{\partial}_s}\fsl{\partial}_{s'}\partial_{s'}^{2(k-1)}\mathcal{S}_{n-2} \nonumber\\
    & &&\hspace{-1.25cm}+\frac{1}{2}i^{-1/2}\partial_s^{2k}\overline{\Psi}_n\overleftarrow{\fsl{\partial}_s}\partial_{s'}^{2k}\mathcal{S}_{n-1} - \frac{1}{2}i^{-1/2}\partial_s^{2k}\overline{\Psi}_{n-1}\fsl{\partial}_{s'}\partial_{s'}^{2k}\mathcal{S}_n \nonumber\\
    & && \hspace{-2.25cm}+i^{1/2}k\partial_s^{2(k-1)}\overline{\Psi}_{n-2}\overleftarrow{\fsl{\partial}_s}\partial_{s'}^{2k}\mathcal{S}_{n-1}-i^{1/2}k\partial_s^{2k}\overline{\Psi}_{n-1}\fsl{\partial}_{s'}\partial_{s'}^{2(k-1)}\mathcal{S}_{n-2} \nonumber \\
     & && \hspace{3.375cm}+ \Big(\frac{1}{2}-k\Big)\partial_s^{2k}\overline{\Psi}_{n-1}\partial_{s'}^{2k}\mathcal{S}_{n-1} \Bigg)\Bigg\}
\end{align}
That this action is gauge invariant is guaranteed by the fact that the field strengths themselves are gauge invariant $\delta\mathcal{S}_{n-i}=0$, and that they satisfy a fermionic massive Bianchi identity
\begin{align}
\label{eq:halfBianch}
\fsl{\partial}_s(\fsl{\partial}_X-m)\mathcal{S}_n-s\cdot\partial_X\partial_s^2\mathcal{S}_n-i^{1/2}(\fsl{\partial}_X-m)\mathcal{S}_{n-1} + i\,s\cdot\partial_X\mathcal{S}_{n-2} = 0
\end{align}
As before, the gauge variation of the action $\delta S_{n+1/2}$ is a straightforward linear combination the left hand side of (\ref{eq:halfBianch}) and the $\delta \mathcal{S}_{n-i}$'s. 
\section{Interactions}
\label{sec:inter}
In this section, we show how to add electromagnetic and gravitational interactions to the actions (\ref{eq:mhyperaction}) and (\ref{eq:mhalfhyperaction}), in a way that preserves the massive gauge symmetry. We begin first with the integer spin case, and then repeat the analysis for half integer spins.
\subsection{Integer spins}
\label{sec:integer}
Naively, when modelling a massive integer spin particle interacting with electromagnetism or gravity, one might try what works for spin $0$, i.e. replace every spacetime derivative $\partial_{X\mu}$ in the action with a covariant derivative $\nabla_{\mu}$, as well as replace every $\eta_{\mu\nu}$ with $g_{\mu\nu}(X)$ and $d^dX$ with $d^dX\sqrt{-g}$ in the case of gravitational interactions. This is often referred to as minimal coupling. Notice however that even this procedure is ambiguous for spins $n\geq 1$, because one can always re-order the derivatives and write $\partial_{X\mu}\partial_{X\nu} = \partial_{X\mu}\partial_{X\nu} + a[\partial_{X\mu},\partial_{X\nu}]$ before converting them to covariant derivatives for any $a$ without changing the free action. Spin $0$ is special in this respect because the only appearance of derivatives in the action is in $\eta^{\mu\nu}\partial_{X\mu}\partial_{X\nu}$. Nevertheless, for the sake of the discussion we will refer to minimal coupling as replacing derivatives with covariant ones in the particular order they appear in (\ref{eq:mhyperaction}).

\sm

A few words must be said about lifting the hyperspace measure (\ref{eq:hyperprod}) to a curved spacetime manifold $M$. Apart from the spacetime integration, this measure is just a formal way to implement index contraction in terms of hyperfields. Within the integration, one should think of the auxiliary vectors as living on the tangent space $s,s'\in T_XM$. One simply integrates over all components of the tangent vectors $s,\,s'$ at the point $X$, weighted by $e^{is\cdot s'}$, where $s\cdot s' = g(s,s') = g_{\mu\nu}s^{\mu}s'^{\nu}$.

\sm

The covariant derivative $\nabla_{\mu}$ on a charge 1 integer spin $n$ hyperfield~$\Phi_n(X,s)$, including both a $U(1)$ gauge field $A_{\mu}$, and a Levi-Civita connection $\Gamma^{\lambda}_{\;\;\mu\nu}$, which implements the standard covariant derivative on its component field $\phi_{\mu_1\cdots\mu_n}(X)$ is 
\begin{align}
    \nabla_{\mu} = \partial_{X\mu} - iA_{\mu} - s^{\nu}\Gamma^{\lambda}_{\;\;\mu\nu}\partial_{s\lambda}
\end{align}
Covariant derivatives no longer commute, but instead their commutator when acting on $\Phi_n(X,s)$ equals
\begin{align}
    [\nabla_{\mu},\nabla_{\nu}] = -iF_{\mu\nu} - s^{\sigma}R^{\omega}_{\;\;\sigma\mu\nu}\partial_{s\omega}
\end{align}
As part of this minimal coupling procedure, we will also deform the massive gauge symmetry (\ref{eq:Stueck}) by replacing spacetime derivatives with covariant derivatives, and $\eta^{\mu\nu}$ with $g^{\mu\nu}$, so that it is consistent with $U(1)$ electromagnetic gauge invariance and general coordinate invariance
\begin{align}
\label{eq:covStueck}
    &\delta\Phi_n = i^{-1/2} s\cdot\nabla \epsilon_{n-1} && \delta\Phi_{n-1} = i^{-1/2} s\cdot \nabla \epsilon_{n-2} + im\,\epsilon_{n-1} \nonumber\\
    &\delta\Phi_{n-2} = -i^{1/2} s\cdot\nabla\partial_s^2\epsilon_{n-1} +2im\,\epsilon_{n-2} && \delta\Phi_{n-3} = -i^{1/2} s\cdot\nabla\partial_s^2\epsilon_{n-2} +3m\partial_s^2\epsilon_{n-1}
\end{align}

\sm

It is straightforward to show that the action (\ref{eq:mhyperaction}) is not invariant under the gauge symmetry (\ref{eq:covStueck}) after minimal coupling, and hence the minimally coupled theory does not consistently describe a massive spin $n$ particle. Indeed, the gauge variation of the minimally coupled action $\delta S_n$ is a linear combination of terms proportional to $\delta\mathcal{F}_{n-i}$, and the covariantized version of the massive Bianchi identities (\ref{eq:intBianch1}), (\ref{eq:intBianch2}). This is an important fact for the subsequent analysis, which follows from the massive action being derived from the corresponding massless action $S_{n,0}$ in $d + 1$ dimensions built from a rank $n$ double traceless hyperfield $\Phi_{n,0}(X,s)$ and massless field strength $\mathcal{F}_{n,0}(X,s)$, described in \cite{Lindwasser:2023zwo}. Indeed, under the gauge transformation $\delta\Phi_{n,0}(X,s) = i^{-1/2}s\cdot\partial_X\epsilon_{n-1}(X,s)$, for $\epsilon_{n-1}(X,s)$ a rank $n-1$ traceless hyperfield, the gauge variation of $S_{n,0}$ is
\begin{align}
    &\delta S_{n,0} = \frac{1}{2}n!\int d^{d+1}X\frac{d^{d+1}sd^{d+1}s'}{(2\pi)^{d+1}}e^{is\cdot s'} \times \nonumber \\
    &\Big(\Phi_{n,0}(X,s)\big(1-\frac{1}{4}s'^2\partial_{s'}^2\big)\delta\mathcal{F}_{n,0}(X,s') - i^{1/2}\epsilon_{n-1}(X,s)\big(\partial_{s'}\cdot\partial_X\mathcal{F}_{n,0} - \frac{1}{2}s'\cdot\partial_X\partial_{s'}^2\mathcal{F}_{n,0}\big)\Big)
\end{align}
After dimensional reduction, the gauge variation of the massless $d+1$ dimensional field strength $\delta\mathcal{F}_{n,0}$ decomposes into the massive $d$ dimensional variations $\delta\mathcal{F}_{n-i}$, and the expression $\partial_{s}\cdot\partial_X\mathcal{F}_{n,0} - \frac{1}{2}s\cdot\partial_X\partial_{s}^2\mathcal{F}_{n,0}$ decomposes into the two massive Bianchi identities (\ref{eq:intBianch1}), (\ref{eq:intBianch2}). These are no longer zero after minimal coupling, but instead equal
\begin{align}
\label{eq:gaugeviol1}
    \delta\mathcal{F}_n &= i^{-1/2}(\nabla^2s\cdot\nabla-s\cdot\nabla[\partial_s\cdot\nabla,s\cdot\nabla])\epsilon_{n-1} \\
    \delta\mathcal{F}_{n-1} &= i^{-1/2}(\nabla^2s\cdot\nabla-s\cdot\nabla[\partial_s\cdot\nabla,s\cdot\nabla])\epsilon_{n-2} + im(\nabla^2-[\partial_s\cdot\nabla,s\cdot\nabla])\epsilon_{n-1} \\
    \delta\mathcal{F}_{n-2} &= -i^{1/2}(\nabla^2s\cdot\nabla-s\cdot\nabla[\partial_s\cdot\nabla,s\cdot\nabla])\partial_s^2\epsilon_{n-1} + 2im(\nabla^2-[\partial_s\cdot\nabla,s\cdot\nabla])\epsilon_{n-2} \\
    \delta\mathcal{F}_{n-3} &= -i^{1/2}(\nabla^2s\cdot\nabla-s\cdot\nabla[\partial_s\cdot\nabla,s\cdot\nabla])\partial_s^2\epsilon_{n-2} + 3m(\nabla^2-[\partial_s\cdot\nabla,s\cdot\nabla])\partial_s^2\epsilon_{n-1} 
\end{align}
\begin{align}
    \partial_s\cdot\nabla&\mathcal{F}_n-\frac{1}{2}s\cdot\nabla\partial_s^2\mathcal{F}_n + \frac{i}{2}s\cdot\nabla\mathcal{F}_{n-2} + i^{1/2}m\mathcal{F}_{n-1} = \nonumber\\
    &(\partial_s\cdot\nabla-\frac{1}{2}s\cdot\nabla\partial_s^2)(\nabla^2-[\partial_s\cdot\nabla,s\cdot\nabla])\Phi_n + [ \partial_s\cdot\nabla-\frac{1}{2}s\cdot\nabla\partial_s^2,[\partial_s\cdot\nabla,s\cdot\nabla]]\Phi_n \nonumber\\
    & \hspace{1cm}+\frac{i}{2}s\cdot\nabla(\nabla^2-[\partial_s\cdot\nabla,s\cdot\nabla])\Phi_{n-2}+\frac{i}{2}[s\cdot\nabla,[\partial_s\cdot\nabla,s\cdot\nabla]]\Phi_{n-2} \nonumber \\
    & \hspace{3cm}+ i^{1/2}m(\nabla^2-[\partial_s\cdot\nabla,s\cdot\nabla])\Phi_{n-1} 
\end{align}
\begin{align}
\label{eq:gaugeviol6}
    \partial_s\cdot\nabla&\mathcal{F}_{n-1} - \frac{1}{2}s\cdot\nabla\partial_s^2\mathcal{F}_{n-1} + \frac{i}{2}s\cdot\nabla\mathcal{F}_{n-3} +\frac{1}{2}i^{-1/2}m\partial_s^2\mathcal{F}_n +\frac{1}{2}i^{1/2}m\mathcal{F}_{n-2}  =  \nonumber \\
     &(\partial_s\cdot\nabla-\frac{1}{2}s\cdot\nabla\partial_s^2)(\nabla^2-[\partial_s\cdot\nabla,s\cdot\nabla])\Phi_{n-1} + [ \partial_s\cdot\nabla-\frac{1}{2}s\cdot\nabla\partial_s^2,[\partial_s\cdot\nabla,s\cdot\nabla]]\Phi_{n-1} \nonumber\\
    & \hspace{1cm}+\frac{i}{2}s\cdot\nabla(\nabla^2-[\partial_s\cdot\nabla,s\cdot\nabla])\Phi_{n-3}+\frac{i}{2}[s\cdot\nabla,[\partial_s\cdot\nabla,s\cdot\nabla]]\Phi_{n-3} \nonumber \\
    &\hspace{.5cm} + \frac{1}{2}i^{-1/2}m(\nabla^2-[\partial_s\cdot\nabla,s\cdot\nabla])\partial_s^2\Phi_n + \frac{1}{2}i^{1/2}m(\nabla^2-[\partial_s\cdot\nabla,s\cdot\nabla])\Phi_{n-2}
\end{align}
One can check that these expressions are proportional to electromagnetic and gravitational curvatures $F_{\mu\nu}$, $R_{\omega\sigma\mu\nu}$.
\subsubsection{Restoring gauge symmetry}
\label{sec:intRestore}
In order to restore the massive gauge symmetry, additional interactions with electromagnetism and gravity must be added to the action. Because the gauge violation of the action $\delta S_n \neq 0$ depends on the properties of the field strengths $\mathcal{F}_{n-i}$, we may consider restoring the gauge symmetry by making appropriate modifications of the field strengths $\mathcal{F}_{n-i}\to\mathcal{F}_{n-i}+\Delta\mathcal{F}_{n-i}$, so that they are again gauge invariant and satisfy the massive Bianchi identities.

\sm

For instance, one could choose $\Delta\mathcal{F}_{n-i}^{(0)}=([\partial_s\cdot\nabla,s\cdot\nabla]-\nabla^2)\Phi_{n-i}$, which cancels all gauge violations in (\ref{eq:gaugeviol1})-(\ref{eq:gaugeviol6}) which are $\mathcal{O}(m)$. The new field strengths $\mathcal{F}_{n-i}^{(0)}\equiv\mathcal{F}_{n-i}+\Delta\mathcal{F}_{n-i}^{(0)}$ still have $\mathcal{O}(1)$ in $m$ gauge violations
\begin{align}
\label{eq:F(0)1}
    \delta\mathcal{F}_n^{(0)} &= i^{-1/2}[[\partial_s\cdot\nabla,s\cdot\nabla],s\cdot\nabla]\epsilon_{n-1} \\
    \delta\mathcal{F}_{n-1}^{(0)} &= i^{-1/2}[[\partial_s\cdot\nabla,s\cdot\nabla],s\cdot\nabla]\epsilon_{n-2} \\
    \delta\mathcal{F}_{n-2}^{(0)} &= -i^{1/2}[[\partial_s\cdot\nabla,s\cdot\nabla],s\cdot\nabla]\partial_s^2\epsilon_{n-1} \\
    \delta\mathcal{F}_{n-3}^{(0)} &= -i^{1/2}[[\partial_s\cdot\nabla,s\cdot\nabla],s\cdot\nabla]\partial_s^2\epsilon_{n-2}
\end{align}
\begin{align}
    \partial_s\cdot\nabla&\mathcal{F}_n^{(0)}-\frac{1}{2}s\cdot\nabla\partial_s^2\mathcal{F}_n^{(0)} + \frac{i}{2}s\cdot\nabla\mathcal{F}_{n-2}^{(0)} + i^{1/2}m\mathcal{F}_{n-1}^{(0)} = \nonumber\\
    & [ \partial_s\cdot\nabla-\frac{1}{2}s\cdot\nabla\partial_s^2,[\partial_s\cdot\nabla,s\cdot\nabla]]\Phi_n + \frac{i}{2}[s\cdot\nabla,[\partial_s\cdot\nabla,s\cdot\nabla]]\Phi_{n-2} 
\end{align}
\begin{align}
\label{eq:F(0)6}
    \partial_s\cdot\nabla&\mathcal{F}_{n-1}^{(0)} - \frac{1}{2}s\cdot\nabla\partial_s^2\mathcal{F}_{n-1}^{(0)} + \frac{i}{2}s\cdot\nabla\mathcal{F}_{n-3}^{(0)} +\frac{1}{2}i^{-1/2}m\partial_s^2\mathcal{F}_n^{(0)} +\frac{1}{2}i^{1/2}m\mathcal{F}_{n-2}^{(0)}  =  \nonumber \\
     &[ \partial_s\cdot\nabla-\frac{1}{2}s\cdot\nabla\partial_s^2,[\partial_s\cdot\nabla,s\cdot\nabla]]\Phi_{n-1} + \frac{i}{2}[s\cdot\nabla,[\partial_s\cdot\nabla,s\cdot\nabla]]\Phi_{n-3}
\end{align}
Because of the appearance of $m$ in the gauge transformations (\ref{eq:covStueck}) and massive Bianchi identities, one can further suppress the gauge violations to $\mathcal{O}(1/m)$, cancelling the $\mathcal{O}(1)$ violations, by adding a $\Delta\mathcal{F}_{n-i}^{(1)}$ which is proportional to $1/m$. This can be done indefinitely, generating an infinite series of $\Delta\mathcal{F}_{n-i}^{(N)}$, each proportional to $1/m^N$, for all integers $N\geq 0$, until the action is exactly gauge invariant.

\sm

To prove this, we proceed by induction. Suppose that we have found modified field strengths $\mathcal{F}_{n-i}^{(2k)}$ which suppress all gauge violations to be $\mathcal{O}(1/m^{2k})$, which we write as
\begin{align}
\label{eq:2k gauge n}
    \delta\mathcal{F}_n^{(2k)} &= \frac{1}{m^{2k}}\hat{\mathcal{O}}_{n}^{(2k)}\epsilon_{n-1} \\
\label{eq:2k gauge n-1}
    \delta\mathcal{F}_{n-1}^{(2k)} &= \frac{1}{m^{2k}}\hat{\mathcal{O}}_{n-1}^{(2k)}\epsilon_{n-2} \\
\label{eq:2k gauge n-2}
    \delta\mathcal{F}_{n-2}^{(2k)} &= \frac{1}{m^{2k}}\hat{\mathcal{O}}_{n-2}^{(2k)}\epsilon_{n-1} \\
\label{eq:2k gauge n-3}
    \delta\mathcal{F}_{n-3}^{(2k)} &= \frac{1}{m^{2k}}\hat{\mathcal{O}}_{n-3}^{(2k)}\epsilon_{n-2}
\end{align}
\begin{align}
\label{eq:2k Bianch n}
    \partial_s\cdot\nabla&\mathcal{F}_n^{(2k)}-\frac{1}{2}s\cdot\nabla\partial_s^2\mathcal{F}_n^{(2k)} + \frac{i}{2}s\cdot\nabla\mathcal{F}_{n-2}^{(2k)} + i^{1/2}m\mathcal{F}_{n-1}^{(2k)} =  \frac{1}{m^{2k}}\hat{\mathcal{U}}_{n}^{(2k)}\Phi_n + \frac{1}{m^{2k}}\hat{\mathcal{U}}_{n-2}^{(2k)}\Phi_{n-2} 
\end{align}
\begin{align}
\label{eq:2k Bianch n-1}
    \partial_s\cdot\nabla\mathcal{F}_{n-1}^{(2k)} - \frac{1}{2}s\cdot\nabla\partial_s^2\mathcal{F}_{n-1}^{(2k)} + \frac{i}{2}s\cdot\nabla\mathcal{F}_{n-3}^{(2k)} +\frac{1}{2}i^{-1/2}m\partial_s^2&\mathcal{F}_n^{(2k)} +\frac{1}{2}i^{1/2}m\mathcal{F}_{n-2}^{(2k)}  =  \nonumber \\
    &\frac{1}{m^{2k}}\hat{\mathcal{U}}_{n-1}^{(2k)}\Phi_{n-1} + \frac{1}{m^{2k}}\hat{\mathcal{U}}_{n-3}^{(2k)}\Phi_{n-3} 
\end{align}
for some operators $\hat{\mathcal{O}}_{n-i}^{(2k)}$, $\hat{\mathcal{U}}_{n-i}^{(2k)}$, where the subscript $n-i$ does not denote the rank, but the hyperfields $\mathcal{F}_{n-i}^{(2k)}$, $\Phi_{n-i}$ they are associated with, respectively. We first show that if $\hat{\mathcal{O}}_{n-i}^{(2k)}$ and $\hat{\mathcal{U}}_{n-i}^{(2k)}$ satisfy conditions (to be determined) ensuring that $\Delta\mathcal{F}_{n-i}^{(2k+1)}$ and $\Delta\mathcal{F}_{n-i}^{(2k+2)}$ can be constructed so that $\mathcal{F}_{n-i}^{(2k+2)} \equiv \mathcal{F}_{n-i}^{(2k)} + \Delta\mathcal{F}_{n-i}^{(2k+1)} + \Delta\mathcal{F}_{n-i}^{(2k+2)}$ suppresses the gauge violations to $\mathcal{O}(1/m^{2k+2})$, then the corresponding operators at the next order $\hat{\mathcal{O}}_{n-i}^{(2k+2)}$ and $\hat{\mathcal{U}}_{n-i}^{(2k+2)}$ will also satisfy those conditions.

\sm

To cancel the $\mathcal{O}(1/m^{2k})$ gauge violations (\ref{eq:2k gauge n}), (\ref{eq:2k gauge n-3}), (\ref{eq:2k Bianch n}) and (\ref{eq:2k Bianch n-1}), one may choose $\Delta\mathcal{F}_{n-i}^{(2k+1)}$ to be
\begin{align}
\label{eq:2k+1 n}
    \Delta\mathcal{F}_{n}^{(2k+1)} &= \frac{i}{m^{2k+1}}\hat{\mathcal{O}}_{n}^{(2k)}\Phi_{n-1} \\
\label{eq:2k+1 n-1}
     \Delta\mathcal{F}_{n-1}^{(2k+1)} &= - \frac{i^{-1/2}}{m^{2k+1}}\Big(\hat{\mathcal{U}}_{n}^{(2k)}\Phi_n + \hat{\mathcal{U}}_{n-2}^{(2k)}\Phi_{n-2}\Big) \\
\label{eq:2k+1 n-2}
     \Delta\mathcal{F}_{n-2}^{(2k+1)} &=- \frac{i^{-1/2}}{m^{2k+1}}\Big((i^{1/2}\partial_s^2\hat{\mathcal{O}}_n^{(2k)}+2\,\hat{\mathcal{U}}_{n-1}^{(2k)})\Phi_{n-1} +2\,\hat{\mathcal{U}}_{n-3}^{(2k)}\Phi_{n-3}\Big) \\
\label{eq:2k+1 n-3}
     \Delta\mathcal{F}_{n-3}^{(2k+1)} &=\frac{i}{2m^{2k+1}}\hat{\mathcal{O}}_{n-3}^{(2k)}\Phi_{n-2}
\end{align}
The remaining $\mathcal{O}(1/m^{2k})$ gauge violations (\ref{eq:2k gauge n-1}) and (\ref{eq:2k gauge n-2}) must then be cancelled by $\delta(\Delta\mathcal{F}_{n-1}^{(2k+1)})$ and $\delta(\Delta\mathcal{F}_{n-2}^{(2k+1)})$, respectively. This in turn is only possible if the following conditions hold
\begin{align}
\label{eq:cond1}
     &\hat{\mathcal{C}}_{1}^{(2k)} \equiv \hat{\mathcal{O}}_{n-1}^{(2k)} - 2i^{1/2}\hat{\mathcal{U}}_{n-2}^{(2k)} = 0 \\
\label{eq:cond2}
    &\hat{\mathcal{C}}_{2}^{(2k)} \equiv \hat{\mathcal{O}}_{n-2}^{(2k)} -i\partial_s^2\hat{\mathcal{O}}_n^{(2k)} - 2i^{1/2}\hat{\mathcal{U}}_{n-1}^{(2k)} - 6i^{-1/2}\hat{\mathcal{U}}_{n-3}^{(2k)}\partial_s^2 = 0
\end{align}
Let us assume this is true and proceed with constructing $\Delta\mathcal{F}_{n-i}^{(2k+2)}$. The gauge violations associated with $\mathcal{F}_{n-i}^{(2k+1)}=\mathcal{F}_{n-i}^{(2k)}+\Delta\mathcal{F}_{n-i}^{(2k+1)}$ are
\begin{align}
\label{eq:2k+1 gauge n}
    \delta\mathcal{F}_n^{(2k+1)} &= \frac{i^{1/2}}{m^{2k+1}}\hat{\mathcal{O}}_{n}^{(2k)}s\cdot\nabla\epsilon_{n-2} \\
\label{eq:2k+1 gauge n-1}
    \delta\mathcal{F}_{n-1}^{(2k+1)} &= \frac{1}{m^{2k+1}}\Big(i\,\hat{\mathcal{U}}_{n}^{(2k)}s\cdot\nabla +\hat{\mathcal{U}}_{n-2}^{(2k)}s\cdot\nabla\partial_s^2\Big)\epsilon_{n-1} \\
\label{eq:2k+1 gauge n-2}
    \delta\mathcal{F}_{n-2}^{(2k+1)} &= \frac{1}{m^{2k+1}}\Big(-i^{-1/2}\partial_s^2\hat{\mathcal{O}}_n^{(2k)}s\cdot\nabla+2i\,\hat{\mathcal{U}}_{n-1}^{(2k)}s\cdot\nabla +2\,\hat{\mathcal{U}}_{n-3}^{(2k)}s\cdot\nabla\partial_s^2\Big)\epsilon_{n-2}\\
\label{eq:2k+1 gauge n-3}
    \delta\mathcal{F}_{n-3}^{(2k+1)} &= \frac{i^{-1/2}}{2m^{2k+1}}\hat{\mathcal{O}}_{n-3}^{(2k)}s\cdot\nabla\partial_s^2\epsilon_{n-1}
\end{align}
\begin{align}
\label{eq:2k+1 Bianch n}
    \partial_s\cdot\nabla&\mathcal{F}_n^{(2k+1)}-\frac{1}{2}s\cdot\nabla\partial_s^2\mathcal{F}_n^{(2k+1)} + \frac{i}{2}s\cdot\nabla\mathcal{F}_{n-2}^{(2k+1)} + i^{1/2}m\mathcal{F}_{n-1}^{(2k+1)} =  \nonumber \\
    &\frac{1}{m^{2k+1}}\Big(i(\partial_s\cdot\nabla-s\cdot\nabla\partial_s^2)\hat{\mathcal{O}}_{n}^{(2k)}-i^{1/2}s\cdot\nabla\,\hat{\mathcal{U}}_{n-1}^{(2k)}\Big)\Phi_{n-1} - \frac{i^{1/2}}{m^{2k+1}}s\cdot\nabla\,\hat{\mathcal{U}}_{n-3}^{(2k)}\Phi_{n-3}
\end{align}
\begin{align}
\label{eq:2k+1 Bianch n-1}
    \partial_s\cdot\nabla&\mathcal{F}_{n-1}^{(2k+1)} - \frac{1}{2}s\cdot\nabla\partial_s^2\mathcal{F}_{n-1}^{(2k+1)} + \frac{i}{2}s\cdot\nabla\mathcal{F}_{n-3}^{(2k+1)} +\frac{1}{2}i^{-1/2}m\partial_s^2\mathcal{F}_n^{(2k+1)} +\frac{1}{2}i^{1/2}m\mathcal{F}_{n-2}^{(2k+1)}  =  \nonumber \\
    &-\frac{i^{-1/2}}{m^{2k+1}}(\partial_s\cdot\nabla-\frac{1}{2}s\cdot\nabla\partial_s^2)(\hat{\mathcal{U}}_{n}^{(2k)}\Phi_{n}+\hat{\mathcal{U}}_{n-2}^{(2k)}\Phi_{n-2}) - \frac{1}{4m^{2k+1}}s\cdot\nabla\hat{\mathcal{O}}_{n-3}^{(2k)}\Phi_{n-2} 
\end{align}
Again, it is straightforward to find modifications $\Delta\mathcal{F}_{n-i}^{(2k+2)}$ which cancel the $\mathcal{O}(1/m^{2k+1})$ gauge violations in (\ref{eq:2k+1 gauge n}), (\ref{eq:2k+1 gauge n-3}), (\ref{eq:2k+1 Bianch n}) and (\ref{eq:2k+1 Bianch n-1})
\begin{align}
\label{eq:2k+2 n}
    \Delta\mathcal{F}_{n}^{(2k+2)} = &-\frac{i^{-1/2}}{2m^{2k+2}}\hat{\mathcal{O}}_{n}^{(2k)}s\cdot\nabla\Phi_{n-2} \\
\label{eq:2k+2 n-1}
     \Delta\mathcal{F}_{n-1}^{(2k+2)} = &- \frac{i^{1/2}}{m^{2k+2}}\Big((\partial_s\cdot\nabla - s\cdot\nabla\partial_s^2)\hat{\mathcal{O}}_{n}^{(2k)}-i^{-1/2}s\cdot\nabla\hat{\mathcal{U}}_{n-1}^{(2k)}\Big)\Phi_{n-1} \nonumber \\
     &+ \frac{1}{m^{2k+2}}s\cdot\nabla\hat{\mathcal{U}}_{n-3}^{(2k)}\Phi_{n-3} \\
\label{eq:2k+2 n-2}
     \Delta\mathcal{F}_{n-2}^{(2k+2)} =& -2\frac{i}{m^{2k+2}}(\partial_s\cdot\nabla-\frac{1}{2}s\cdot\nabla\partial_s^2)(\hat{\mathcal{U}}_{n}^{(2k)}\Phi_{n} +\hat{\mathcal{U}}_{n-2}^{(2k)}\Phi_{n-2}) \nonumber \\
     &+\frac{i^{-1/2}}{2m^{2k+2}}(s\cdot\nabla\hat{\mathcal{O}}_{n-3}^{(2k)}-i\partial_s^2\hat{\mathcal{O}}_{n}^{(2k)}s\cdot\nabla)\Phi_{n-2} \\
\label{eq:2k+2 n-3}
     \Delta\mathcal{F}_{n-3}^{(2k+2)} =&\frac{i^{1/2}}{2m^{2k+2}}\hat{\mathcal{O}}_{n-3}^{(2k)}s\cdot\nabla\partial_s^2\Phi_{n-1}
\end{align}
Demanding that $\delta(\Delta\mathcal{F}_{n-1}^{(2k+2)})$ and $\delta(\Delta\mathcal{F}_{n-2}^{(2k+2)})$ also cancel the $\mathcal{O}(1/m^{2k+1})$ gauge violations in (\ref{eq:2k+1 gauge n-1}) and (\ref{eq:2k+1 gauge n-2}) imposes further conditions on $\hat{\mathcal{O}}_{n-i}^{(2k)}$ and $\hat{\mathcal{U}}_{n-i}^{(2k)}$
\begin{align}
\label{eq:cond3}
    \hat{\mathcal{C}}_{3}^{(2k)}\equiv\,& i^{-1/2}s\cdot\nabla\hat{\mathcal{O}}_{n-3}^{(2k)} + 2\,\hat{\mathcal{U}}_{n-1}^{(2k)}s\cdot\nabla \nonumber \\
    &-4i(\partial_s\cdot\nabla - \frac{1}{2}s\cdot\nabla\partial_s^2)\hat{\mathcal{U}}_{n-2}^{(2k)} - 2i\,\hat{\mathcal{U}}_{n-3}^{(2k)}s\cdot \nabla\partial_s^2 = 0\\
\label{eq:cond4}
    \hat{\mathcal{C}}_{4}^{(2k)}\equiv \,&i^{1/2}(\partial_s\cdot\nabla-s\cdot\nabla\partial_s^2)\hat{\mathcal{O}}_n^{(2k)} -\hat{\mathcal{U}}_n^{(2k)}s\cdot\nabla \nonumber \\
    & - s\cdot\nabla\,\hat{\mathcal{U}}_{n-1}^{(2k)} + i\,\hat{\mathcal{U}}_{n-2}^{(2k)}s\cdot\nabla\partial_s^2 + 3is\cdot\nabla\,\hat{\mathcal{U}}_{n-3}^{(2k)}\partial_s^2 = 0
\end{align}
Assuming that these conditions are satisfied, we may use $\mathcal{F}_{n-i}^{(2k+2)} = \mathcal{F}_{n-i}^{(2k+1)} + \Delta\mathcal{F}_{n-i}^{(2k+2)}$ to find the corresponding operators at the next order $\hat{\mathcal{O}}_{n-i}^{(2k+2)}$ and $\hat{\mathcal{U}}_{n-i}^{(2k+2)}$ associated with $\mathcal{O}(1/m^{2k+2})$ gauge violations
\begin{align}
    \hat{\mathcal{O}}_{n}^{(2k+2)} =& \frac{1}{2}\hat{\mathcal{O}}_{n}^{(2k)}(s\cdot\nabla)^2\partial_s^2 \\ 
    \hat{\mathcal{O}}_{n-1}^{(2k+2)} =& -(\partial_s\cdot\nabla - s\cdot\nabla\partial_s^2)\hat{\mathcal{O}}_{n}^{(2k)}s\cdot\nabla \nonumber \\
    &+ i^{-1/2}s\cdot\nabla\,\hat{\mathcal{U}}_{n-1}^{(2k)} s\cdot\nabla -i^{1/2}s\cdot\nabla\,\hat{\mathcal{U}}_{n-3}^{(2k)} s\cdot\nabla\partial_s^2 \\
    \hat{\mathcal{O}}_{n-2}^{(2k+2)} =& -2i^{1/2}(\partial_s\cdot\nabla - \frac{1}{2}s\cdot\nabla\partial_s^2)(\hat{\mathcal{U}}_{n}^{(2k)}s\cdot\nabla -i\,\hat{\mathcal{U}}_{n-2}^{(2k)}s\cdot\nabla\partial_s^2) \nonumber \\
    &+ \frac{i}{2}\partial_s^2\hat{\mathcal{O}}_{n}^{(2k)}(s\cdot\nabla)^2\partial_s^2 - \frac{1}{2}s\cdot\nabla \hat{\mathcal{O}}_{n-3}^{(2k)} s\cdot\nabla\partial_s^2 \\
    \hat{\mathcal{O}}_{n-3}^{(2k+2)} =& \,\frac{1}{2}\hat{\mathcal{O}}_{n-3}^{(2k)}s\cdot\nabla\partial_s^2 s\cdot\nabla 
\end{align}
\begin{align}
    \hat{\mathcal{U}}_{n}^{(2k+2)} =& \,s\cdot\nabla (\partial_s\cdot\nabla - \frac{1}{2}s\cdot\nabla\partial_s^2)\hat{\mathcal{U}}_{n}^{(2k)} \\
    \hat{\mathcal{U}}_{n-1}^{(2k+2)} =& -i^{1/2}(\partial_s\cdot\nabla - \frac{1}{2}s\cdot\nabla\partial_s^2)(\partial_s\cdot\nabla - s\cdot\nabla\partial_s^2)\hat{\mathcal{O}}_{n}^{(2k)} \nonumber \\
    &+ (\partial_s\cdot\nabla - \frac{1}{2}s\cdot\nabla\partial_s^2)s\cdot\nabla\hat{\mathcal{U}}_{n-1}^{(2k)}-\frac{1}{4}i^{-1/2}s\cdot\nabla \hat{\mathcal{O}}_{n-3}^{(2k)} s\cdot\nabla\partial_s^2 \\
    \hat{\mathcal{U}}_{n-2}^{(2k+2)} =& \,s\cdot\nabla (\partial_s\cdot\nabla - \frac{1}{2}s\cdot\nabla\partial_s^2)\hat{\mathcal{U}}_{n-2}^{(2k)} + \frac{1}{4}i^{1/2}(s\cdot\nabla)^2\hat{\mathcal{O}}_{n-3}^{(2k)} \nonumber \\
    &- \frac{1}{2}i^{-1/2}(\partial_s\cdot\nabla - s\cdot\nabla\partial_s^2)\hat{\mathcal{O}}_{n}^{(2k)}s\cdot\nabla \\
    \hat{\mathcal{U}}_{n-3}^{(2k+2)} =& (\partial_s\cdot\nabla - \frac{1}{2}s\cdot\nabla\partial_s^2)s\cdot\nabla\,\hat{\mathcal{U}}_{n-3}^{(2k)}
\end{align}

\sm

To recapitulate, given the existence of field strengths $\mathcal{F}_{n-i}^{(2k)}$ that satisfy (\ref{eq:2k gauge n}) -- (\ref{eq:2k Bianch n-1}), we have found four conditions (\ref{eq:cond1}), (\ref{eq:cond2}), (\ref{eq:cond3}), and (\ref{eq:cond4}) on the operators $\hat{\mathcal{O}}_{n-i}^{(2k)}$ and $\hat{\mathcal{U}}_{n-i}^{(2k)}$, which if satisfied enables the construction of $\Delta\mathcal{F}_{n-i}^{(2k+1)}$ and $\Delta\mathcal{F}_{n-i}^{(2k+2)}$, resulting in improved field strengths $\mathcal{F}_{n-i}^{(2k+2)} = \mathcal{F}_{n-i}^{(2k)} + \Delta\mathcal{F}_{n-i}^{(2k+1)} + \Delta\mathcal{F}_{n-i}^{(2k+2)}$ with $\mathcal{O}(1/m^{2k+2})$ gauge violations. It is worth mentioning at this stage that this procedure is not unique. For instance, one could add a term proportional $\Phi_n$ to $\Delta\mathcal{F}_{n-3}^{(2k+1)}$. This would add to the action a term linear in $\Phi_n$, and so we ignore this possibility.

\sm

The next step is to show that if $\hat{\mathcal{O}}_{n-i}^{(2k)}$ and $\hat{\mathcal{U}}_{n-i}^{(2k)}$ satisfy $\hat{\mathcal{C}}_j^{(2k)} = 0$ for $j = 1,2,3,4$, then $\hat{\mathcal{O}}_{n-i}^{(2k+2)}$ and $\hat{\mathcal{U}}_{n-i}^{(2k+2)}$ satisfy $\hat{\mathcal{C}}_j^{(2k+2)} = 0$. This follows after writing $\hat{\mathcal{O}}_{n-i}^{(2k+2)}$ and $\hat{\mathcal{U}}_{n-i}^{(2k+2)}$ in terms of $\hat{\mathcal{O}}_{n-i}^{(2k)}$ and $\hat{\mathcal{U}}_{n-i}^{(2k)}$
\begin{align}
    \hat{\mathcal{C}}_1^{(2k+2)} = &\,\frac{1}{2}i^{-1/2}s\cdot\nabla \hat{\mathcal{C}}_3^{(2k)} = 0 \\ 
    \hat{\mathcal{C}}_2^{(2k+2)} = & \,2 i^{1/2} (\partial_s\cdot\nabla - \frac{1}{2}s\cdot\nabla\partial_s^2)\hat{\mathcal{C}}_4^{(2k)} = 0 \\
    \hat{\mathcal{C}}_3^{(2k+2)} = & \,(\partial_s\cdot\nabla - \frac{1}{2}s\cdot\nabla\partial_s^2) s\cdot\nabla \hat{\mathcal{C}}_3^{(2k)} =  0 \\
    \hat{\mathcal{C}}_4^{(2k+2)} = & \,s\cdot\nabla(\partial_s\cdot\nabla - \frac{1}{2}s\cdot\nabla\partial_s^2) \hat{\mathcal{C}}_4^{(2k)} =  0
\end{align}

\sm

The final step is to verify that the operators $\hat{\mathcal{O}}_{n-i}^{(0)}$ and $\hat{\mathcal{U}}_{n-i}^{(0)}$ from (\ref{eq:F(0)1}) -- (\ref{eq:F(0)6}) satisfy $\hat{\mathcal{C}}_j^{(0)} = 0$. The operators $\hat{\mathcal{O}}_{n-i}^{(0)}$ and $\hat{\mathcal{U}}_{n-i}^{(0)}$ associated with minimal coupling are
\begin{align}
    &\hat{\mathcal{O}}_{n}^{(0)} = i^{-1/2}[[\partial_s\cdot\nabla,s\cdot\nabla],s\cdot\nabla] && \hat{\mathcal{U}}_{n}^{(0)} = [\partial_s\cdot\nabla - \frac{1}{2}s\cdot\nabla\partial_s^2,[\partial_s\cdot\nabla,s\cdot\nabla]] \\ 
    &\hat{\mathcal{O}}_{n-1}^{(0)} = i^{-1/2}[[\partial_s\cdot\nabla,s\cdot\nabla],s\cdot\nabla] && \hat{\mathcal{U}}_{n-1}^{(0)} = [\partial_s\cdot\nabla - \frac{1}{2}s\cdot\nabla\partial_s^2,[\partial_s\cdot\nabla,s\cdot\nabla]] \\
    &\hat{\mathcal{O}}_{n-2}^{(0)} = -i^{1/2}[[\partial_s\cdot\nabla,s\cdot\nabla],s\cdot\nabla]\partial_s^2 && \hat{\mathcal{U}}_{n-2}^{(0)} = \frac{i}{2}[s\cdot\nabla,[\partial_s\cdot\nabla,s\cdot\nabla]] \\
     &\hat{\mathcal{O}}_{n-3}^{(0)} = -i^{1/2}[[\partial_s\cdot\nabla,s\cdot\nabla],s\cdot\nabla]\partial_s^2 && \hat{\mathcal{U}}_{n-3}^{(0)} = \frac{i}{2}[s\cdot\nabla,[\partial_s\cdot\nabla,s\cdot\nabla]]
\end{align}
That these satisfy $\hat{\mathcal{C}}_j^{(0)} = 0$ is straightforward to verify, thus providing the base case for the induction argument to go through. The modifications $\Delta\mathcal{F}_{n-i}^{(N)}$ (\ref{eq:2k+1 n}) -- (\ref{eq:2k+1 n-3}) and (\ref{eq:2k+2 n}) -- (\ref{eq:2k+2 n-3}) provide a gauge invariant completion of minimal coupling.
\subsubsection{Non-minimal gauge invariant interactions}
\label{sec:nonmin}
As we saw in the previous section, minimally coupling $\Phi_{n-i}$ to electromagnetism and gravity does not preserve the massive gauge symmetry (\ref{eq:covStueck}), but there exists modifications $\Delta\mathcal{F}_{n-i}^{(N)}$ to the fields strengths $\mathcal{F}_{n-i} \to \mathcal{F}_{n-i} + \sum_{N = 0}^{\infty}\Delta\mathcal{F}_{n-i}^{(N)}$ which restores the gauge symmetry exactly.

\sm

Apart from minimal coupling, which gives $\Phi_n$ a charge/mass, there are non-minimal couplings one might consider adding to your theory. For instance, one might further consider adding a term $\Delta\mathcal{F}_{n}^{(0)} = g_0([\partial_s\cdot\nabla,s\cdot\nabla]-\nabla^2)\Phi_n$ to $\mathcal{F}_n$, which is related to the magnetic dipole and gravitational quadrupole moment couplings, because
\begin{align}
\label{eq:intdiquad}
    [\partial_s\cdot\nabla,s\cdot\nabla]-\nabla^2 = i s^{\mu}F_{\mu\nu}\partial_s^{\nu} + s^{\mu}R_{\mu\nu}\partial_s^{\nu} - s^{\sigma}s^{\nu}R_{\omega\sigma\mu\nu}\partial_s^{\mu}\partial_s^{\omega}
\end{align}
and, as explained in \cite{Cortese:2013lda}, this is precisely how such a coupling occurs at the level of the equations of motion. Adding such terms will, like in the case of minimal coupling, break massive gauge invariance. Luckily, the analysis of \autoref{sec:intRestore} provides a simple prescription for adding further interactions which restores the gauge symmetry. Indeed, we may consider interactions seeded by an \textit{arbitrary} non-minimal coupling operator $\hat{\mathcal{M}}^{(2k)}$ with mass dimension $2k+2$ which is invariant with respect to the massive gauge symmetry, and adding 
\begin{align}
    &\Delta \mathcal{F}_n^{(2k)} = \frac{1}{m^{2k}}\hat{\mathcal{M}}^{(2k)}\Phi_n, && \Delta \mathcal{F}_{n-2}^{(2k)} = \frac{i}{m^{2k}}\partial_s^2\hat{\mathcal{M}}^{(2k)}\Phi_n
\end{align}
This introduces $\mathcal{O}(1/m^{2k})$ gauge violations, with associated operators $\hat{\mathcal{O}}_{n-i}^{(2k)}$ and $\hat{\mathcal{U}}_{n-i}^{(2k)}$ equal to 
\begin{align}
    \hat{\mathcal{O}}_{n}^{(2k)} &= i^{-1/2}\hat{\mathcal{M}}^{(2k)}s\cdot\nabla \\
    \hat{\mathcal{O}}_{n-2}^{(2k)} &= i^{1/2}\partial_s^2\hat{\mathcal{M}}^{(2k)}s\cdot\nabla \\
    \hat{\mathcal{U}}_{n}^{(2k)} &= (\partial_s\cdot\nabla - s\cdot\nabla\partial_s^2)\hat{\mathcal{M}}^{(2k)}
\end{align}
while the rest are zero. These operators trivially satisfy the four conditions $\hat{\mathcal{C}}_j^{(2k)} = 0$ for any $\hat{\mathcal{M}}^{(2k)}$. Therefore, the gauge symmetry can be restored in exactly the same manner described in \autoref{sec:intRestore}. Multipole and tidal couplings are obviously accommodated by this class of gauge invariant interactions.

\subsubsection{Resummed non-local field strengths}
\label{sec:intnonlocal}
Gauge invariance requires infinitely many terms in the action which are arbitrarily high order in derivatives, making interactions inherently non-local in nature. To make this manifest, we may formally resum the modifications $\Delta\mathcal{F}_{n-i}^{(N)}$ and write the interactions in terms of non-local operators.

\sm

Suppose we have field strengths $\mathcal{F}_{n-i}^{(2k)}$ with $\mathcal{O}(1/m^{2k})$ gauge violations that can be cancelled via modifications $\Delta\mathcal{F}_{n-i}^{(N)}$ for all $N > 2k$ in the way described in \autoref{sec:intRestore}. The gauge invariant field strengths $\mathcal{F}_{n-i} = \mathcal{F}_{n-i}^{(2k)} + \sum_{N = 2k +1}^{\infty}\Delta\mathcal{F}_{n-i}^{(N)}$ may then be written as
\begin{align}
\label{eq:ginvF1}
    &\mathcal{F}_{n} = \mathcal{F}_n^{(2k)} +\frac{i}{m^{2k+1}}\hat{\mathcal{O}}_{n}^{(2k)}\frac{1}{1-\frac{1}{2m^2}(s\cdot\nabla)^2\partial_s^2}X_{n-1}\\
    &\mathcal{F}_{n-1} = \mathcal{F}_{n-1}^{(2k)}-\frac{i^{-1/2}}{m^{2k+1}}\frac{1}{1-\frac{1}{m^2}s\cdot\nabla(\partial_s\cdot\nabla - \frac{1}{2}s\cdot\nabla\partial_s^2)}\Big(\hat{\mathcal{U}}_n^{(2k)}\Phi_n + \hat{\mathcal{U}}_{n-2}^{(2k)}\Phi_{n-2}\Big) \nonumber \\
    &+\frac{1}{m^{2k+2}}\frac{1}{1-\frac{1}{m^2}s\cdot\nabla(\partial_s\cdot\nabla - \frac{1}{2}s\cdot\nabla\partial_s^2)}s\cdot\nabla\Big(\hat{\mathcal{U}}_{n-1}^{(2k)}\Phi_{n-1}+\hat{\mathcal{U}}_{n-3}^{(2k)}\Phi_{n-3}\Big) \nonumber \\
    &-\frac{i^{1/2}}{m^{2k+2}}\frac{1}{1-\frac{1}{m^2}s\cdot\nabla(\partial_s\cdot\nabla - \frac{1}{2}s\cdot\nabla\partial_s^2)}(\partial_s\cdot\nabla - s\cdot\nabla\partial_s^2)\hat{\mathcal{O}}_n^{(2k)}\frac{1}{1-\frac{1}{2m^2}(s\cdot\nabla)^2\partial_s^2}X_{n-1}\nonumber\\
    &-\frac{1}{4m^{2k+3}}\frac{1}{1-\frac{1}{m^2}s\cdot\nabla(\partial_s\cdot\nabla - \frac{1}{2}s\cdot\nabla\partial_s^2)}(s\cdot\nabla)^2\hat{\mathcal{O}}_{n-3}^{(2k)}\frac{1}{1-\frac{1}{2m^2}s\cdot\nabla\partial_s^2s\cdot\nabla}X_{n-2} \\
    &\mathcal{F}_{n-2} = \mathcal{F}_{n-2}^{(2k)}-2\frac{i^{-1/2}}{m^{2k+1}}\frac{1}{1-\frac{1}{m^2}(\partial_s\cdot\nabla - \frac{1}{2}s\cdot\nabla\partial_s^2)s\cdot\nabla} \Big(\hat{\mathcal{U}}_{n-1}^{(2k)}\Phi_{n-1} +\hat{\mathcal{U}}_{n-3}^{(2k)}\Phi_{n-3}\Big) \nonumber\\
    &-\frac{1}{m^{2k+1}}\frac{1}{1-\frac{1}{m^2}(\partial_s\cdot\nabla - \frac{1}{2}s\cdot\nabla\partial_s^2)s\cdot\nabla}\partial_s^2\hat{\mathcal{O}}_n^{(2k)}\frac{1}{1-\frac{1}{2m^2}(s\cdot\nabla)^2\partial_s^2}X_{n-1} \nonumber\\
    &-2\frac{i}{m^{2k+2}}\frac{1}{1-\frac{1}{m^2}(\partial_s\cdot\nabla - \frac{1}{2}s\cdot\nabla\partial_s^2)s\cdot\nabla}(\partial_s\cdot\nabla - \frac{1}{2}s\cdot\nabla\partial_s^2)\Big(\hat{\mathcal{U}}_n^{(2k)}\Phi_n+\hat{\mathcal{U}}_{n-2}^{(2k)}\Phi_{n-2}\Big)\nonumber\\
    &+\frac{i^{-1/2}}{2m^{2k+2}}\frac{1}{1-\frac{1}{m^2}(\partial_s\cdot\nabla - \frac{1}{2}s\cdot\nabla\partial_s^2)s\cdot\nabla}s\cdot\nabla\hat{\mathcal{O}}_{n-3}^{(2k)}\frac{1}{1-\frac{1}{2m^2}s\cdot\nabla\partial_s^2s\cdot\nabla}X_{n-2} \nonumber\\
    &+\frac{2}{m^{2k+3}}\frac{1}{1-\frac{1}{m^2}(\partial_s\cdot\nabla - \frac{1}{2}s\cdot\nabla\partial_s^2)s\cdot\nabla}(\partial_s\cdot\nabla-\frac{1}{2}s\cdot\nabla\partial_s^2)^2\hat{\mathcal{O}}_n^{(2k)}\frac{1}{1-\frac{1}{2m^2}(s\cdot\nabla)^2\partial_s^2}X_{n-1} \\
    \label{eq:ginvF4}
    &\mathcal{F}_{n-3} = \mathcal{F}_{n-3}^{(2k)} + \frac{i}{2m^{2k+1}}\hat{\mathcal{O}}_{n-3}^{(2k)}\frac{1}{1-\frac{1}{2m^2}s\cdot\nabla\partial_s^2s\cdot\nabla}X_{n-2}
\end{align}
where we have defined $X_{n-1}\equiv\Phi_{n-1}+\frac{i^{1/2}}{2m}s\cdot\nabla\Phi_{n-2}$ and $X_{n-2}\equiv\Phi_{n-2}+\frac{i^{-1/2}}{m}s\cdot\nabla\partial_s^2\Phi_{n-1}$. One can check that these expressions are gauge invariant and satisfy the covariantized massive Bianchi identities. To simplify these expressions, we are free to gauge fix, by for instance choosing $\Phi_{n-1},\Phi_{n-2}=0$
\begin{align}
    \mathcal{F}_{n} &= \mathcal{F}_n^{(2k)} \\
    \mathcal{F}_{n-1} &= \mathcal{F}_{n-1}^{(2k)}-\frac{i^{-1/2}}{m^{2k+1}}\frac{1}{1-\frac{1}{m^2}s\cdot\nabla(\partial_s\cdot\nabla - \frac{1}{2}s\cdot\nabla\partial_s^2)}\hat{\mathcal{U}}_n^{(2k)}\Phi_n \nonumber \\
    &\phantom{=}+\frac{1}{m^{2k+2}}\frac{1}{1-\frac{1}{m^2}s\cdot\nabla(\partial_s\cdot\nabla - \frac{1}{2}s\cdot\nabla\partial_s^2)}s\cdot\nabla\hat{\mathcal{U}}_{n-3}^{(2k)}\Phi_{n-3}\\
    \mathcal{F}_{n-2} &= \mathcal{F}_{n-2}^{(2k)}-2\frac{i^{-1/2}}{m^{2k+1}}\frac{1}{1-\frac{1}{m^2}(\partial_s\cdot\nabla - \frac{1}{2}s\cdot\nabla\partial_s^2)s\cdot\nabla} \hat{\mathcal{U}}_{n-3}^{(2k)}\Phi_{n-3} \nonumber\\
    &\phantom{=}-2\frac{i}{m^{2k+2}}\frac{1}{1-\frac{1}{m^2}(\partial_s\cdot\nabla - \frac{1}{2}s\cdot\nabla\partial_s^2)s\cdot\nabla}(\partial_s\cdot\nabla - \frac{1}{2}s\cdot\nabla\partial_s^2)\hat{\mathcal{U}}_n^{(2k)}\Phi_n\\
    \mathcal{F}_{n-3} &= \mathcal{F}_{n-3}^{(2k)}
\end{align}

\sm

For spins $n=0,1$, the non-local gauge invariant completions truncate to a local operator, so that a local field theory is possible. For $n\geq 2$ on the other hand, non-locality appears to be unavoidable, without performing field redefinitions.
\subsubsection{Interactions linear in spin $n$ fields}
\label{sec:linearint}
If the massive spin $n$ particle has $U(1)$ charge 0, there is another class of interactions, i.e. those which are only linear in the matter fields $\Phi_{n-i}$ in the action, which can be made gauge invariant. To get a charge 0 particle, it is sufficient to impose the reality conditions on the fields $\tilde{\Phi}_n=\Phi_n$, $\tilde{\Phi}_{n-1}=-\Phi_{n-1}$, $\tilde{\Phi}_{n-2}=\Phi_{n-2}$, and $\tilde{\Phi}_{n-3}=-\Phi_{n-3}$. The gauge parameters in turn are constrained $\tilde{\epsilon}_{n-1} = \epsilon_{n-1}$, $\tilde{\epsilon}_{n-2} = -\epsilon_{n-2}$. If we rewrite everything in terms of real fields by performing the field redefinitions $\Phi_{n-1}\to i\Phi_{n-1}$, $\Phi_{n-3}\to i\Phi_{n-3}$, and $\epsilon_{n-2}\to i\epsilon_{n-2}$, the covariantized massive gauge transformations become
\begin{align}
\label{eq:nuetralcovStueck}
    &\delta\Phi_n = i^{-1/2} s\cdot\nabla \epsilon_{n-1} && \delta\Phi_{n-1} = i^{-1/2} s\cdot \nabla \epsilon_{n-2} + m\,\epsilon_{n-1} \nonumber\\
    &\delta\Phi_{n-2} = -i^{1/2} s\cdot\nabla\partial_s^2\epsilon_{n-1} - 2m\,\epsilon_{n-2} && \delta\Phi_{n-3} = -i^{1/2} s\cdot\nabla\partial_s^2\epsilon_{n-2} - 3im\partial_s^2\epsilon_{n-1}
\end{align}
where $\nabla_{\mu}=\partial_{X\mu}-s^{\nu}\Gamma^{\lambda}_{\;\;\mu\nu}\partial_{s\lambda}$.

\sm

Coupling $\Phi_{n-i}$ to hyperfields $\mathcal{J}_{n-i}$ which are gauge invariant functions of the gravitational and possibly electromagnetic fields by adding to the action
\begin{align}
\label{eq:lineartemplate}
    \Delta S_n = n!\int d^dX\frac{d^dsd^ds'}{(2\pi)^d}e^{is\cdot s'}\Phi_{n-i}(X,s)\mathcal{J}_{n-i}(X,s')
\end{align}
where the sum over $i=0,\dots,3$ is implied, is gauge invariant provided that the $\mathcal{J}_{n-i}$ satisfy
\begin{align}
\label{eq:linear1}
    &\partial_s\cdot\nabla\mathcal{J}_n + is^2\partial_s\cdot\nabla\mathcal{J}_{n-2} - i^{-1/2}m\mathcal{J}_{n-1} - 3i^{1/2}ms^2\mathcal{J}_{n-3}=0\\
\label{eq:linear2}
    &\partial_s\cdot\nabla\mathcal{J}_{n-1} + is^2\partial_s\cdot\nabla\mathcal{J}_{n-3} + 2i^{-1/2}m\mathcal{J}_{n-2} =0
\end{align}

\sm

To construct $\mathcal{J}_{n-i}$ which satisfy (\ref{eq:linear1}) and (\ref{eq:linear2}), we may proceed in a similar fashion to \autoref{sec:intRestore}, by starting with $\mathcal{J}_{n-i}$ which do not satisfy (\ref{eq:linear1}) and (\ref{eq:linear2}), and finding improvement terms inductively that make the interactions exactly gauge invariant. Instead, we may more simply search for gauge invariant linear combinations of the fields $\Phi_{n-i}$. There are two gauge invariant linear combinations
\begin{align}
    &A_n \equiv \Phi_n-\frac{i^{-1/2}}{m}s\cdot\nabla\frac{1}{1-\frac{1}{2m^2}(s\cdot\nabla)^2\partial_s^2}X_{n-1} \\
    &A_{n-3} \equiv \Phi_{n-3} + 3i\partial_s^2\frac{1}{1-\frac{1}{2m^2}(s\cdot\nabla)^2\partial_s^2}X_{n-1}-\frac{i^{1/2}}{2m}s\cdot\nabla\partial_s^2\frac{1}{1-\frac{1}{2m^2}s\cdot\nabla\partial_s^2s\cdot\nabla}X_{n-2}
\end{align}
where now $X_{n-1}=\Phi_{n-1}+\frac{i^{-1/2}}{2m}s\cdot\nabla\Phi_{n-2}$ and $X_{n-2}=\Phi_{n-2}+\frac{i^{1/2}}{m}s\cdot\nabla\partial_s^2\Phi_{n-1}$. Note that these gauge invariant combinations are non-local for $n\geq 3$. Using $A_n$ and $A_{n-3}$, we may construct gauge invariant interactions linear in $\Phi_{n-i}$ of any type, including the ones classified in \cite{Chakraborty:2020rxf}
\begin{align}
\label{eq:linearinvariant}
    \Delta S_n = n!\int d^dX\frac{d^dsd^ds'}{(2\pi)^d}e^{is\cdot s'}\Big(A_n(X,s)\mathcal{J}_n(X,s') + A_{n-3}(X,s)\mathcal{J}_{n-3}(X,s')\Big)
\end{align}
for any $\mathcal{J}_{n}$ and $\mathcal{J}_{n-3}$. Rewriting (\ref{eq:linearinvariant}) in the form (\ref{eq:lineartemplate}) through integration by parts will give $\mathcal{J}_{n-1}$ and $\mathcal{J}_{n-2}$ that are functions of $\mathcal{J}_{n}$ and $\mathcal{J}_{n-3}$, which altogether satisfy (\ref{eq:linear1}) and (\ref{eq:linear2}).
\subsubsection{Spin 1}
For concreteness, we present here the action for a massive spin 1 particle interacting with electromagnetism and gravity resulting from restoring the gauge symmetry after minimal coupling in \autoref{sec:intRestore}. For spin 1, all non-local operators present in the general case truncate to local ones. The gauge invariant field strengths $\mathcal{F}_1$ and $\mathcal{F}_0$ take the form
\begin{align}
    \mathcal{F}_1 =& ([\partial_s\cdot\nabla,s\cdot\nabla]-m^2-s\cdot\nabla\partial_s\cdot\nabla)\Phi_1-i^{1/2}ms\cdot\nabla\Phi_0 +\frac{i^{1/2}}{m}[[\partial_s\cdot\nabla,s\cdot\nabla],s\cdot\nabla]\Phi_0\\
    \mathcal{F}_0 =& \nabla^2\Phi_0 + i^{-1/2}m\partial_s\cdot\nabla\Phi_1\nonumber\\
    &-\frac{i^{-1/2}}{m}[\partial_s\cdot\nabla,[\partial_s\cdot\nabla,s\cdot\nabla]]\Phi_1-\frac{1}{m^2}[\partial_s\cdot\nabla,[\partial_s\cdot\nabla,s\cdot\nabla]]s\cdot\nabla\Phi_0
\end{align}
The action is then simply (\ref{eq:mhyperaction}) at $n=1$
\begin{align}
    S_1 = \frac{1}{2}\int d^dX\frac{d^ds d^ds'}{(2\pi)^d}e^{is\cdot s'}\Big(\tilde{\Phi}_1(X,s)\mathcal{F}_1(X,s')+\tilde{\Phi}_0(X,s)\mathcal{F}_0(X,s')+\text{c.c.}\Big)
\end{align}
\subsubsection{Spin 2}
\label{sec:2}
For spin 2, there is also a simplification. Some of the non-local operators truncate to local ones, while others remain non-local, but can be brought into a simpler form. The operators $(1-\frac{1}{2m^2}(s\cdot\nabla)^2\partial_s^2)^{-1}$ and $(1-\frac{1}{2m^2}s\cdot\nabla\partial_s^2s\cdot\nabla)^{-1}$ act respectively on rank 1 and rank 0 hyperfields in (\ref{eq:ginvF1}) -- (\ref{eq:ginvF4}) when $n=2$. These operators act on such hyperfields as the identity operator $1$. The other two non-local operators $(1-\frac{1}{m^2}s\cdot\nabla(\partial_s\cdot\nabla-\frac{1}{2}s\cdot\nabla\partial_s^2))^{-1}$ and $(1-\frac{1}{m^2}(\partial_s\cdot\nabla-\frac{1}{2}s\cdot\nabla\partial_s^2)s\cdot\nabla)^{-1}$ act again respectively on rank 1 and rank 0 hyperfields, but in this case have a non-trivial representation. They may be written manifestly as Hermitian operators with respect to (\ref{eq:hyperprod}) as
\begin{align}
    \frac{1}{1-\frac{1}{m^2}s\cdot\nabla(\partial_s\cdot\nabla-\frac{1}{2}s\cdot\nabla\partial_s^2)} &= 1 + \frac{1}{m^2}s\cdot\nabla\frac{1}{1-\frac{1}{m^2}\nabla^2}\partial_s\cdot\nabla \\
     \frac{1}{1-\frac{1}{m^2}(\partial_s\cdot\nabla-\frac{1}{2}s\cdot\nabla\partial_s^2)s\cdot\nabla} &=\frac{1}{1-\frac{1}{m^2}\nabla^2}
\end{align}
In the case of minimal coupling as in \autoref{sec:intRestore}, the gauge invariant field strengths $\mathcal{F}_2$, $\mathcal{F}_1$, and $\mathcal{F}_0$ take the form
\begin{align}
    \mathcal{F}_2 =&\, ([\partial_s\cdot\nabla,s\cdot\nabla]-m^2-s\cdot\nabla\partial_s\cdot\nabla + \frac{1}{2}(s\cdot\nabla)^2\partial_s^2)\Phi_2 - \frac{i}{2}(s\cdot\nabla)^2\Phi_0 - i^{1/2}ms\cdot\nabla\Phi_1 \nonumber \\
    &+\frac{i^{1/2}}{m}[[\partial_s\cdot\nabla,s\cdot\nabla],s\cdot\nabla](\Phi_1 + \frac{i^{1/2}}{2m}s\cdot\nabla\Phi_0) \\
    \mathcal{F}_1 = &\,([\partial_s\cdot\nabla,]-s\cdot\nabla\partial_s\cdot\nabla)\Phi_1 + i^{-1/2}(\partial_s\cdot\nabla - s\cdot\nabla\partial_s^2)\Phi_2 \nonumber \\
    & -\frac{i^{-1/2}}{m}\Big(1 + \frac{1}{m^2}s\cdot\nabla\frac{1}{1-\frac{1}{m^2}\nabla^2}\partial_s\cdot\nabla\Big)[\partial_s\cdot\nabla-\frac{1}{2}s\cdot\nabla\partial_s^2,[\partial_s\cdot\nabla,s\cdot\nabla]]\Phi_2 \nonumber \\
    &+\frac{1}{m^2}\Big(1 + \frac{1}{m^2}s\cdot\nabla\frac{1}{1-\frac{1}{m^2}\nabla^2}\partial_s\cdot\nabla\Big)(\partial_s\cdot\nabla-\frac{1}{2}s\cdot\nabla\partial_s^2)[s\cdot\nabla,[\partial_s\cdot\nabla,s\cdot\nabla]]\Phi_1 \nonumber \\
    & -\frac{i^{1/2}}{2m}\Big(1 + \frac{1}{m^2}s\cdot\nabla\frac{1}{1-\frac{1}{m^2}\nabla^2}\partial_s\cdot\nabla\Big)[s\cdot\nabla,[\partial_s\cdot\nabla,s\cdot\nabla]])\Phi_0 \nonumber\\
    & +\frac{i^{1/2}}{2m^3}\Big(1 + \frac{1}{m^2}s\cdot\nabla\frac{1}{1-\frac{1}{m^2}\nabla^2}\partial_s\cdot\nabla\Big)(\partial_s\cdot\nabla - s\cdot\nabla\partial_s^2)[s\cdot\nabla,[\partial_s\cdot\nabla,s\cdot\nabla]]s\cdot\nabla\Phi_0 \\
    \mathcal{F}_0 = & \,\nabla^2\Phi_0 - im^2\partial_s^2\Phi_2 + 2i^{-1/2}m\partial_s\cdot\nabla\Phi_1 \nonumber \\
    &-\frac{2i}{m^2}\frac{1}{1-\frac{1}{m^2}\nabla^2}\partial_s\cdot\nabla [\partial_s\cdot\nabla-\frac{1}{2}s\cdot\nabla\partial_s^2,[\partial_s\cdot\nabla,s\cdot\nabla]]\Phi_2 \nonumber \\
    & -\frac{2i^{-1/2}}{m^3}\frac{1}{1-\frac{1}{m^2}\nabla^2}\partial_s\cdot\nabla (\partial_s\cdot\nabla-\frac{1}{2}s\cdot\nabla\partial_s^2) [s\cdot\nabla,[\partial_s\cdot\nabla,s\cdot\nabla]](\Phi_1 + \frac{i^{1/2}}{2m}s\cdot\nabla\Phi_0)
\end{align}
The action is then (\ref{eq:mhyperaction}) at $n=2$
\begin{align}
    S_2=\int d^dX\frac{d^dsd^ds'}{(2\pi)^d}e^{is\cdot s'}\Big(&\tilde{\Phi}_2\mathcal{F}_2+\frac{1}{4}\partial_s^2\tilde{\Phi}_2\partial_{s'}^2\mathcal{F}_2-\frac{i}{4}\partial_s^2\tilde{\Phi}_2\mathcal{F}_0 \nonumber \\
    &-\frac{i}{4}\tilde{\Phi}_0\partial_{s'}^2\mathcal{F}_2+\frac{1}{4}\tilde{\Phi}_0\mathcal{F}_0+\tilde{\Phi}_1\mathcal{F}_1 + \text{c.c.}\Big)
\end{align}
\subsection{Half integer spins}
\label{sec:halfinteger}
In this section, we repeat the analysis of \autoref{sec:integer} for massive half integer spin particles described by the free action (\ref{eq:mhalfhyperaction}). We will find gauge invariant interactions through a completely analogous argument.

\sm

We first begin by minimally coupling our theory to electromagnetism and gravity. This amounts to replacing every spacetime derivative $\partial_{X\mu}$ in the action with a covariant derivative $\nabla_{\mu}$, as well as again replacing every $\eta_{\mu\nu}$ with $g_{\mu\nu}(X)$, and $d^dX$ with $d^dX\sqrt{-g}$. In the case of half integer spin fields, we also use Dirac matrices $\gamma_{\mu}(X)$ satisfying~$\{\gamma_{\mu},\gamma_{\nu}\} = 2g_{\mu\nu}$. This is made possible by introducing a vielbein $e_{\mu}^{\;a}(X)$ satisfying $e_{\mu}^{\;a}e_{\nu}^{\;b} \eta_{ab} = g_{\mu\nu}$ and $e_{\mu}^{\;a}e_{\nu}^{\;b} g^{\mu\nu} = \eta^{ab}$, and writing $\gamma_{\mu}(X) = e_{\mu}^{\;a}(X)\gamma_{a}$, where $\gamma_a$ are constant Dirac matrices satisfying $\{\gamma_a,\gamma_b\} = 2\eta_{ab}$. Here Latin indices $a,b = 0,\dots , d-1$ are raised and lowered with the flat metric $\eta^{ab}$ and $\eta_{ab}$, respectively.

\sm

The covariant derivative $\nabla_{\mu}$ on a charge 1 half integer spin $n+1/2$ hyperfield $\Psi_n(X,s)$, including a $U(1)$ gauge field $A_{\mu}$, a spin connection $\omega_{\mu}^{\;ab}$ and Levi-Civita connection $\Gamma^{\lambda}_{\;\;\mu\nu}$ related via $\Gamma^{\lambda}_{\;\;\mu\nu} = e^{\lambda}_{\;a}(\partial_{\mu}e_{\nu}^{\;a}+\omega_{\mu}^{\;ab}e_{\nu b})$, which implements the standard covariant derivative on its component field $\psi_{\mu_1\cdots\mu_n}(X)$ is
\begin{align}
    \nabla_{\mu}=\partial_{X\mu}-iA_{\mu}+\frac{1}{4}\omega_{\mu}^{\;ab}\gamma_{ab}-s^{\nu}\Gamma^{\lambda}_{\;\;\mu\nu}\partial_{s\lambda}
\end{align}
Covariant derivatives no longer commute, but instead their commutator when acting on $\Psi_n(X,s)$ equals
\begin{align}
    [\nabla_{\mu},\nabla_{\nu}] = -iF_{\mu\nu} + \frac{1}{4}R_{\omega\sigma\mu\nu}\gamma^{\omega\sigma}  - s^{\sigma}R^{\omega}_{\;\;\sigma\mu\nu}\partial_{s\omega}
\end{align}
As part of minimal coupling, we will as before deform the massive gauge symmetry (\ref{eq:halfStueck}) by replacing spacetime derivatives with covariant derivatives, $\eta^{\mu\nu}$ with $g^{\mu\nu}$, and $\gamma_{\mu}$ with $e_{\mu}^{\;a}\gamma_a$, so that it is consistent with $U(1)$ electromagnetic gauge invariance and general coordinate invariance
\begin{align}
\label{eq:halfcovStueck}
&\delta\Psi_n=i^{-1/2}s\cdot\nabla\epsilon_{n-1} \nonumber\\
&\delta\Psi_{n-1} = -i s\cdot\nabla\fsl{\partial}_s\epsilon_{n-1}+im\epsilon_{n-1} \nonumber\\
&\delta\Psi_{n-2} = -i^{1/2}s\cdot\nabla\partial_s^2\epsilon_{n-1}+2i^{1/2}m\fsl{\partial}_s\epsilon_{n-1}
\end{align}

\sm

The action (\ref{eq:mhalfhyperaction}) is not invariant under the gauge symmetry (\ref{eq:halfcovStueck}) after minimal coupling, and hence does not consistently describe a massive spin $n + 1/2$ particle. Indeed, the gauge variation of the minimally coupled action $\delta S_{n+1/2}$ is a linear combination of terms proportional to $\delta\mathcal{S}_{n-i}$, and the covariantized version of the fermionic massive Bianchi identity (\ref{eq:halfBianch}). This follows from the fact that the massive action is derived from the corresponding massless action $S_{n+1/2,0}$ in $d + 1$ dimensions built from a rank $n$ triple $\gamma$ traceless hyperfield $\Psi_{n,0}(X,s)$ and massless field strength $\mathcal{S}_{n,0}(X,s)$, described in \cite{Lindwasser:2023zwo}. Under the gauge transformation $\delta\Psi_{n,0}(X,s) = i^{-1/2}s\cdot\partial_X\epsilon_{n-1}(X,s)$, for $\epsilon_{n-1}(X,s)$ a rank $n-1$ $\gamma$ traceless hyperfield, the gauge variation of $S_{n+1/2,0}$ is
\begin{align}
    &\delta S_{n+1/2,0} = -n!\int d^{d+1}X\frac{d^{d+1}sd^{d+1}s'}{(2\pi)^{d+1}}e^{is\cdot s'} \times \nonumber \\
    &\Big(\overline{\Psi}_{n,0}(X,s)\big(1-\frac{1}{2}\fsl{s'}\fsl{\partial}_{s'}-\frac{1}{4}s'^2\partial_{s'}^2\big)\delta\mathcal{S}_{n,0}(X,s') - \frac{1}{2}i^{1/2}\overline{\epsilon}_{n-1}(X,s)\big(\fsl{\partial}_{s'}\fsl{\partial}_X - s'\cdot\partial_X\partial_{s'}^2\big)\mathcal{S}_{n,0}\Big)
\end{align}
After dimensional reduction, the gauge variation of the massless $d+1$ dimensional field strength $\delta\mathcal{S}_{n,0}$ decomposes into the massive $d$ dimensional variations $\delta\mathcal{S}_{n-i}$, and the expression $\fsl{\partial}_{s}\fsl{\partial}_X\mathcal{S}_{n,0} - s\cdot\partial_X\partial_{s}^2\mathcal{S}_{n,0}$ decomposes into the fermionic massive Bianchi identity (\ref{eq:halfBianch}). These are no longer zero after minimal coupling, but instead equal
\begin{align}
\label{eq:halfgaugeviolation1}
    \delta\mathcal{S}_n &= i^{-1/2}[\fsl{\nabla},s\cdot\nabla]\epsilon_{n-1} \\
    \delta\mathcal{S}_{n-1} &= -i[\fsl{\nabla},s\cdot\nabla]\fsl{\partial}_s\epsilon_{n-1} \\
    \delta\mathcal{S}_{n-2} &= -i^{1/2}[\fsl{\nabla},s\cdot\nabla]\partial_s^2\epsilon_{n-1}
\end{align}
\begin{align}
\label{eq:halfgaugeviolation4}
    &\fsl{\partial}_s(\fsl{\nabla} - m)\mathcal{S}_n-s\cdot\nabla\partial_s^2\mathcal{S}_n-i^{1/2}(\fsl{\nabla}-m)\mathcal{S}_{n-1}+i\,s\cdot\nabla\mathcal{S}_{n-2} = \nonumber \\
    \big([\fsl{\partial}_s&,(\fsl{\nabla})^2] - \fsl{\partial}_s[\fsl{\nabla},s\cdot\nabla]\fsl{\partial}_s\big)\Psi_n + i^{1/2}\{\fsl{\partial}_s,[\fsl{\nabla},s\cdot\nabla]\}\Psi_{n-1} -i[\fsl{\nabla},s\cdot\nabla]\Psi_{n-2}
\end{align}
\subsubsection{Restoring gauge symmetry}
\label{sec:halfRestor}
We now show how to restore the gauge symmetry after minimal coupling by adding further interactions with electromagnetism and gravity in the form of making appropriate modifications of the field strengths $\mathcal{S}_{n-i}\to\mathcal{S}_{n-i} + \Delta\mathcal{S}_{n-i}$, so that they are again gauge invariant and satisfy the massive Bianchi identity, in a way exactly analogous to the integer spin case in \autoref{sec:intRestore}. 

\sm

In particular, first note that the gauge violations in (\ref{eq:halfgaugeviolation1}) -- (\ref{eq:halfgaugeviolation4}), are $\mathcal{O}(1)$ in $m$. Because of the appearance of $m$ in the gauge transformations (\ref{eq:halfcovStueck}) and massive Bianchi identity, it is possible to suppress the gauge violations to $\mathcal{O}(1/m)$, cancelling the $\mathcal{O}(1)$ violations, by adding a $\Delta\mathcal{S}_{n-i}^{(1)}$ which is proportional to $1/m$. We will prove by induction that this process can continue indefinitely, generating an infinite series of $\Delta\mathcal{S}_{n-i}^{(N)}$, each proportional to $1/m^N$, for all integers $N\geq 0$, until the action is exactly gauge invariant.

\sm

Suppose we have found modified field strengths $\mathcal{S}_{n-i}^{(2k)}$ which suppress all gauge violations to be $\mathcal{O}(1/m^{2k})$, which we write as
\begin{align}
\label{eq:half 2k gauge viol1}
    \delta\mathcal{S}_{n}^{(2k)} =& \frac{1}{m^{2k}}\hat{\mathcal{O}}_{n}^{(2k)}\epsilon_{n-1} \\
\label{eq:half 2k gauge viol2}
    \delta\mathcal{S}_{n-1}^{(2k)} =& \frac{1}{m^{2k}}\hat{\mathcal{O}}_{n-1}^{(2k)}\epsilon_{n-1} \\
\label{eq:half 2k gauge viol3}
    \delta\mathcal{S}_{n-2}^{(2k)} =& \frac{1}{m^{2k}}\hat{\mathcal{O}}_{n-2}^{(2k)}\epsilon_{n-1} 
\end{align}
\begin{align}
\label{eq:half 2k Bianch }
     \fsl{\partial}_s(\fsl{\nabla} - m)&\mathcal{S}_n^{(2k)}-s\cdot\nabla\partial_s^2\mathcal{S}_n^{(2k)}-i^{1/2}(\fsl{\nabla}-m)\mathcal{S}_{n-1}^{(2k)}+i\,s\cdot\nabla\mathcal{S}_{n-2}^{(2k)} = \nonumber \\
     &\frac{1}{m^{2k}}\hat{\mathcal{U}}_{n}^{(2k)}\Psi_n +\frac{1}{m^{2k}}\hat{\mathcal{U}}_{n-1}^{(2k)}\Psi_{n-1} +\frac{1}{m^{2k}}\hat{\mathcal{U}}_{n-2}^{(2k)}\Psi_{n-2}  
\end{align}
for some operators $\hat{\mathcal{O}}_{n-i}^{(2k)}$, $\hat{\mathcal{U}}_{n-i}^{(2k)}$. We first show that if $\hat{\mathcal{O}}_{n-i}^{(2k)}$ and $\hat{\mathcal{U}}_{n-i}^{(2k)}$ satisfy conditions (to be determined) ensuring that $\Delta\mathcal{S}_{n-i}^{(2k+1)}$ and $\Delta\mathcal{S}_{n-i}^{(2k+2)}$ can be constructed so that $\mathcal{S}_{n-i}^{(2k+2)} \equiv \mathcal{S}_{n-i}^{(2k)} + \Delta\mathcal{S}_{n-i}^{(2k+1)} + \Delta\mathcal{S}_{n-i}^{(2k+2)}$ suppresses the gauge violations to $\mathcal{O}(1/m^{2k+2})$, then the corresponding operators at the next order $\hat{\mathcal{O}}_{n-i}^{(2k+2)}$ and $\hat{\mathcal{U}}_{n-i}^{(2k+2)}$ will also satisfy those conditions.

\sm

To cancel the $\mathcal{O}(1/m^{2k})$ gauge violations (\ref{eq:half 2k gauge viol1}), (\ref{eq:half 2k gauge viol3}), and (\ref{eq:half 2k Bianch }), one may choose $\Delta\mathcal{S}_{n-i}^{(2k+1)}$ to be
\begin{align}
\label{eq:half 2k+1 n}
    \Delta\mathcal{S}_{n}^{(2k+1)} &= \frac{i}{m^{2k+1}}\hat{\mathcal{O}}_{n}^{(2k)}\Psi_{n-1} \\
\label{eq:half 2k+1 n-1}
    \Delta\mathcal{S}_{n-1}^{(2k+1)} &= -\frac{i^{-1/2}}{m^{2k+1}}\Big(\hat{\mathcal{U}}_{n}^{(2k)}\Psi_n +(-i\fsl{\partial}_s\hat{\mathcal{O}}_n^{(2k)}+\hat{\mathcal{U}}_{n-1}^{(2k)})\Psi_{n-1} +\hat{\mathcal{U}}_{n-2}^{(2k)}\Psi_{n-2}\Big) \\
\label{eq:half 2k+1 n-2}
    \Delta\mathcal{S}_{n-2}^{(2k+1)} &= \frac{i}{m^{2k+1}}\hat{\mathcal{O}}_{n-2}^{(2k)}\Psi_{n-1}
\end{align}
The remaining $\mathcal{O}(1/m^{2k})$ gauge violation (\ref{eq:half 2k gauge viol2}) must then be cancelled by $\delta(\Delta\mathcal{S}_{n-1}^{(2k+1)})$. This in turn is only possible if the following condition holds
\begin{align}
   \hat{\mathcal{C}}_1^{(2k)} \equiv \hat{\mathcal{O}}_{n-1}^{(2k)} -i^{-1/2}\fsl{\partial}_s\hat{\mathcal{O}}_n^{(2k)} - i^{1/2}\hat{\mathcal{U}}_{n-1}^{(2k)} -2\,\hat{\mathcal{U}}_{n-2}^{(2k)}\fsl{\partial}_s = 0
\end{align}
Let us assume this is true and proceed with constructing $\Delta\mathcal{S}_{n-i}^{(2k+2)}$. The gauge violations associated with $\mathcal{S}_{n-i}^{(2k+1)}=\mathcal{S}_{n-i}^{(2k)} + \Delta\mathcal{S}_{n-i}^{(2k+1)}$ are
\begin{align}
\label{eq:half 2k+1 gaugeviol1}
    \delta\mathcal{S}_{n}^{(2k+1)} &= \frac{1}{m^{2k+1}}\hat{\mathcal{O}}_{n}^{(2k)}s\cdot\nabla\fsl{\partial}_s\epsilon_{n-1}\\
\label{eq:half 2k+1 gaugeviol2}
    \delta\mathcal{S}_{n-1}^{(2k+1)} &= \frac{1}{m^{2k+1}}\Big(i^{1/2}(-i\fsl{\partial}_s\hat{\mathcal{O}}_{n}^{(2k)} + \hat{\mathcal{U}}_{n-1}^{(2k)})s\cdot\nabla\fsl{\partial}_s + i\,\hat{\mathcal{U}}_{n}^{(2k)}s\cdot\nabla + \hat{\mathcal{U}}_{n-2}^{(2k)}s\cdot\nabla \partial_s^2\Big)\epsilon_{n-1} \\
\label{eq:half 2k+1 gaugeviol3}
    \delta\mathcal{S}_{n-2}^{(2k+1)} &= \frac{1}{m^{2k+1}}\hat{\mathcal{O}}_{n-2}^{(2k)}s\cdot\nabla\fsl{\partial}_s\epsilon_{n-1}
\end{align}
\begin{align}
\label{eq:half 2k +1 Bianch }
     \fsl{\partial}_s(\fsl{\nabla}& - m)\mathcal{S}_n^{(2k+1)}-s\cdot\nabla\partial_s^2\mathcal{S}_n^{(2k+1)}-i^{1/2}(\fsl{\nabla}-m)\mathcal{S}_{n-1}^{(2k+1)}+i\,s\cdot\nabla\mathcal{S}_{n-2}^{(2k+1)} = \nonumber \\
      &\frac{1}{m^{2k+1}}\Big(\fsl{\nabla}\hat{\mathcal{U}}_{n-1}^{(2k)} + i([\fsl{\partial}_s,\fsl{\nabla}] - s\cdot\nabla\partial_s^2)\hat{\mathcal{O}}_n^{(2k)}- s\cdot\nabla \hat{\mathcal{O}}_{n-2}^{(2k)}\Big)\Psi_{n-1} \nonumber \\
     &\hspace{1.5cm}+\frac{1}{m^{2k+1}}\fsl{\nabla}\hat{\mathcal{U}}_{n}^{(2k)}\Psi_n +\frac{1}{m^{2k+1}}\fsl{\nabla}\hat{\mathcal{U}}_{n-2}^{(2k)}\Psi_{n-2} 
\end{align}
It is straightforward to find $\Delta\mathcal{S}_{n-i}^{(2k+2)}$ which cancel the $\mathcal{O}(1/m^{2k+1})$ gauge violations in (\ref{eq:half 2k+1 gaugeviol1}), (\ref{eq:half 2k+1 gaugeviol3}), and (\ref{eq:half 2k +1 Bianch })
\begin{align}
\label{eq:half 2k+2 n}
    &\Delta\mathcal{S}_{n}^{(2k+2)} = \frac{i}{m^{2k+2}}\hat{\mathcal{O}}_{n}^{(2k)}s\cdot\nabla\fsl{\partial}_s\Psi_{n-1} \\
\label{eq:half 2k+2 n-1}
    &\Delta\mathcal{S}_{n-1}^{(2k+2)} = -\frac{i^{-1/2}}{m^{2k+2}}\Big( \fsl{\nabla}\hat{\mathcal{U}}_{n}^{(2k)}\Psi_n +\fsl{\nabla}\hat{\mathcal{U}}_{n-2}^{(2k)}\Psi_{n-2} \Big)\nonumber \\
    &-\frac{i^{-1/2}}{m^{2k+2}}\Big(
    \fsl{\nabla}\hat{\mathcal{U}}_{n-1}^{(2k)} -i\fsl{\partial}_s\hat{\mathcal{O}}_{n}^{(2k)}s\cdot\nabla\fsl{\partial}_s + i([\fsl{\partial}_s,\fsl{\nabla}] - s\cdot\nabla\partial_s^2)\hat{\mathcal{O}}_n^{(2k)}- s\cdot\nabla \hat{\mathcal{O}}_{n-2}^{(2k)}\Big)\Psi_{n-1}\\
\label{eq:half 2k+2 n-2}
    &\Delta\mathcal{S}_{n-2}^{(2k+2)} = \frac{i}{m^{2k+2}}\hat{\mathcal{O}}_{n-2}^{(2k)}s\cdot\nabla\fsl{\partial}_s\Psi_{n-1} 
\end{align}
Demanding that $\delta(\Delta\mathcal{S}_{n-1}^{(2k+2)})$ also cancels the remaining $\mathcal{O}(1/m^{2k+1})$ gauge violation in (\ref{eq:half 2k+1 gaugeviol2}) imposes one final condition on $\hat{\mathcal{O}}_{n-i}^{(2k)}$ and $\hat{\mathcal{U}}_{n-i}^{(2k)}$
\begin{align}
    \hat{\mathcal{C}}_2^{(2k)}\equiv&\, i^{1/2}([\fsl{\partial}_s,\fsl{\nabla}] - s\cdot\nabla\partial_s^2)\hat{\mathcal{O}}_{n}^{(2k)} - i^{-1/2}s\cdot\nabla\hat{\mathcal{O}}_{n-2}^{(2k)}-\hat{\mathcal{U}}_{n}^{(2k)}s\cdot\nabla \nonumber \\
    &+i^{-1/2}\fsl{\nabla}\hat{\mathcal{U}}_{n-1}^{(2k)} - i^{-1/2}\hat{\mathcal{U}}_{n-1}^{(2k)}s\cdot\nabla\fsl{\partial}_s-2i\fsl{\nabla}\hat{\mathcal{U}}_{n-2}^{(2k)}\fsl{\partial}_s + i\,\hat{\mathcal{U}}_{n-2}^{(2k)}s\cdot\nabla\partial_s^2 =0
\end{align}
Assuming that this condition is satisfied, we may use $\mathcal{S}_{n-i}^{(2k+2)} = \mathcal{S}_{n-i}^{(2k+1)} + \Delta\mathcal{S}_{n-i}^{(2k+2)}$ to find the corresponding operators at the next order $\hat{\mathcal{O}}_{n-i}^{(2k+2)}$ and $\hat{\mathcal{U}}_{n-i}^{(2k+2)}$ associated with $\mathcal{O}(1/m^{2k+2})$ gauge violations
\begin{align}
    \hat{\mathcal{O}}_n^{(2k+2)} =& \,\hat{\mathcal{O}}_n^{(2k)}(s\cdot\nabla\fsl{\partial}_s)^2 \\
    \hat{\mathcal{O}}_{n-1}^{(2k+2)} =& \,i^{-1/2}\fsl{\partial}_s\hat{\mathcal{O}}_n^{(2k)}(s\cdot\nabla\fsl{\partial}_s)^2 -i^{-1/2}([\fsl{\partial}_s,\fsl{\nabla}] - s\cdot\nabla\partial_s^2)\hat{\mathcal{O}}_{n}^{(2k)}s\cdot\nabla\fsl{\partial}_s + i\fsl{\nabla}\hat{\mathcal{U}}_{n}^{(2k)}s\cdot\nabla \nonumber\\
    & +i^{1/2}\fsl{\nabla}\hat{\mathcal{U}}_{n-1}^{(2k)}s\cdot\nabla\fsl{\partial}_s + \fsl{\nabla}\hat{\mathcal{U}}_{n-2}^{(2k)}s\cdot\nabla\partial_s^2 - i^{1/2}s\cdot\nabla\hat{\mathcal{O}}_{n-2}^{(2k)}s\cdot\nabla\fsl{\partial}_s \\
    \hat{\mathcal{O}}_{n-2}^{(2k+2)} =&\, \hat{\mathcal{O}}_{n-2}^{(2k)}(s\cdot\nabla\fsl{\partial}_s)^2 \\
    \hat{\mathcal{U}}_n^{(2k+2)} =& \, (\fsl{\nabla})^2\hat{\mathcal{U}}_n^{(2k)} \\
    \hat{\mathcal{U}}_{n-1}^{(2k+2)} =& \,  i([\fsl{\partial}_s,\fsl{\nabla}] - s\cdot\nabla\partial_s^2)\hat{\mathcal{O}}_{n}^{(2k)}s\cdot\nabla\fsl{\partial}_s + i\fsl{\nabla}([\fsl{\partial}_s,\fsl{\nabla}] - s\cdot\nabla\partial_s^2)\hat{\mathcal{O}}_{n}^{(2k)} \nonumber\\
    & +(\fsl{\nabla})^2\hat{\mathcal{U}}_{n-1}^{(2k)}- \fsl{\nabla}s\cdot\nabla\hat{\mathcal{O}}_{n-2}^{(2k)}-s\cdot\nabla\hat{\mathcal{O}}_{n-2}^{(2k)}s\cdot\nabla\fsl{\partial}_s\\
    \hat{\mathcal{U}}_{n-2}^{(2k+2)} =& \, (\fsl{\nabla})^2\hat{\mathcal{U}}_{n-2}^{(2k)}
\end{align}

\sm

The next step is to show that if $\hat{\mathcal{O}}_{n-i}^{(2k)}$ and $\hat{\mathcal{U}}_{n-i}^{(2k)}$ satisfy $\hat{\mathcal{C}}_{j}^{(2k)}=0$ for $j=1,2$, then $\hat{\mathcal{O}}_{n-i}^{(2k+2)}$ and $\hat{\mathcal{U}}_{n-i}^{(2k+2)}$ satisfy $\hat{\mathcal{C}}_{j}^{(2k+2)}=0$. This follows after writing $\hat{\mathcal{O}}_{n-i}^{(2k+2)}$ and $\hat{\mathcal{U}}_{n-i}^{(2k+2)}$ in terms of $\hat{\mathcal{O}}_{n-i}^{(2k)}$ and $\hat{\mathcal{U}}_{n-i}^{(2k)}$
\begin{align}
    \hat{\mathcal{C}}_1^{(2k+2)} &= -i\fsl{\nabla}\hat{\mathcal{C}}_2^{(2k)} = 0 \\
    \hat{\mathcal{C}}_2^{(2k+2)} &= (\fsl{\nabla})^2\hat{\mathcal{C}}_2^{(2k)} = 0
\end{align}

\sm

Finally, we must verify that the operators $\hat{\mathcal{O}}_{n-i}^{(0)}$ and $\hat{\mathcal{U}}_{n-i}^{(0)}$ associated with minimal coupling (\ref{eq:halfgaugeviolation1}) -- (\ref{eq:halfgaugeviolation4}) satisfy $\hat{\mathcal{C}}_j^{(0)}=0$. $\hat{\mathcal{O}}_{n-i}^{(0)}$ and $\hat{\mathcal{U}}_{n-i}^{(0)}$ are equal to
\begin{align}
    &\hat{\mathcal{O}}_{n}^{(0)} = i^{-1/2}[\fsl{\nabla},s\cdot\nabla], &&\hat{\mathcal{U}}_{n}^{(0)} = [\fsl{\partial}_s,(\fsl{\nabla})^2] -\fsl{\partial}_s[\fsl{\nabla},s\cdot\nabla]\fsl{\partial}_s \\
    &\hat{\mathcal{O}}_{n-1}^{(0)} = -i[\fsl{\nabla},s\cdot\nabla]\fsl{\partial}_s, && \hat{\mathcal{U}}_{n-1}^{(0)} = i^{1/2}\{\fsl{\partial}_s,[\fsl{\nabla},s\cdot\nabla]\}  \\
    &\hat{\mathcal{O}}_{n-2}^{(0)} = -i^{1/2}[\fsl{\nabla},s\cdot\nabla]\partial_s^2, &&
    \hat{\mathcal{U}}_{n-2}^{(0)} =-i[\fsl{\nabla},s\cdot\nabla]
\end{align}
That these satisfy $\hat{\mathcal{C}}_j^{(0)} = 0$ is easily verified, thus providing the base case for the induction argument to go through. The modifications $\Delta\mathcal{S}_{n-i}^{(N)}$ (\ref{eq:half 2k+1 n}) -- (\ref{eq:half 2k+1 n-2}) and (\ref{eq:half 2k+2 n}) -- (\ref{eq:half 2k+2 n-2}) provide a gauge invariant completion of minimal coupling.
\subsubsection{Non-minimal gauge invariant interactions}
\label{sec:halfnonmin}
As we saw in the previous section, minimally coupling $\Psi_{n-i}$ to electromagnetism and gravity does not preserve the massive gauge symmetry (\ref{eq:halfcovStueck}), but there exists modifications $\Delta\mathcal{S}_{n-i} \to \mathcal{S}_{n-i} + \sum_{N = 0}^{\infty}\Delta\mathcal{S}_{n-i}^{(N)}$ which restores the gauge symmetry exactly.

\sm

We would also like to know how to add non-minimal couplings to electromagnetism and gravity in a gauge invariant way. For instance, one might consider adding a term $\Delta\mathcal{S}_n^{(1)} = \frac{1}{m}g_1([\fsl{\partial}_s\fsl{\nabla},s\cdot\nabla]-(\fsl{\nabla})^2)\Psi_n$ to $\mathcal{S}_n$, which is related to the magnetic dipole and gravitational quadrupole moment couplings, because it is linear in curvatures with no spacetime derivatives
\begin{align}
    [\fsl{\partial}_s\fsl{\nabla},s\cdot\nabla]-(\fsl{\nabla})^2 = \fsl{\partial}_s[\fsl{\nabla},s\cdot\nabla] &= is^{\mu}F_{\mu\nu}\partial_s^{\nu} + i\gamma^{\mu\nu}F_{\mu\nu} - is^{\mu}F_{\mu\nu}\gamma^{\nu\lambda}\partial_{s\lambda} \nonumber\\
    &+ s^{\mu}R_{\mu\nu}\partial_s^{\nu} - s^{\sigma}s^{\nu}R_{\omega\sigma\mu\nu}\partial_s^{\mu}\partial_s^{\omega} \nonumber\\ 
    &- \gamma^{\mu\nu}s^{\omega}R_{\omega\sigma\mu\nu}\partial_s^{\sigma} - \gamma^{\sigma\mu}s^{\nu}R_{\omega\sigma\mu\nu}\partial_s^{\omega} + s^{\sigma}s^{\nu}R_{\omega\sigma\mu\nu}\gamma^{\mu\lambda}\partial_{s\lambda}\partial_s^{\omega}
\end{align}
Adding such a term will break massive gauge invariance. Luckily there is, again, a prescription for adding further interactions which restores the gauge invariance in exactly the same manner described in \autoref{sec:halfRestor}. We show two classes of non-minimal interactions which can be made gauge invariant. The first is seeded by the following non-minimal coupling modifications, associated with an arbitrary operator $\hat{\mathcal{M}}^{(2k+1)}$ with mass dimension $2k+2$ which is invariant with respect to the massive gauge symmetry 
\begin{align}
    &\Delta\mathcal{S}_n^{(2k+1)} = \frac{1}{m^{2k+1}}\hat{\mathcal{M}}^{(2k+1)}\Psi_n \\ & \Delta\mathcal{S}_{n-1}^{(2k+1)} = \frac{i^{-1/2}}{m^{2k+1}}\fsl{\partial}_s\hat{\mathcal{M}}^{(2k+1)}\Psi_n \\
    &\Delta\mathcal{S}_n^{(2k+2)} = \frac{i^{1/2}}{m^{2k+2}}\hat{\mathcal{M}}^{(2k+1)}s\cdot\nabla\Psi_{n-1}, \\& \Delta\mathcal{S}_{n-1}^{(2k+2)} = -\frac{i^{-1/2}}{m^{2k+2}}\Big(([\fsl{\partial}_s,\fsl{\nabla}]-s\cdot\nabla\partial_s^2)\hat{\mathcal{M}}^{(2k+1)}\Psi_n-i^{1/2}\fsl{\partial}_s\hat{\mathcal{M}}^{(2k+1)}s\cdot\nabla\Psi_{n-1}\Big)
\end{align}
These modifications introduce $\mathcal{O}(1/m^{2k+2})$ gauge violations, described by the operators $\hat{\mathcal{O}}_{n-i}^{(2k+2)}$ and $\hat{\mathcal{U}}_{n-i}^{(2k+2)}$ equal to
\begin{align}
    &\hat{\mathcal{O}}_{n}^{(2k+2)} = i^{-1/2}\hat{\mathcal{M}}^{(2k+1)}(s\cdot\nabla)^2\fsl{\partial}_s \\
    &\hat{\mathcal{O}}_{n-1}^{(2k+2)} = i([\fsl{\partial}_s,\fsl{\nabla}]-s\cdot\nabla\partial_s^2)\hat{\mathcal{M}}^{(2k+1)}s\cdot\nabla - i\fsl{\partial}_s\hat{\mathcal{M}}^{(2k+1)}(s\cdot\nabla)^2\fsl{\partial}_s \\
    &\hat{\mathcal{U}}_{n}^{(2k+2)} = \fsl{\nabla}([\fsl{\partial}_s,\fsl{\nabla}]-s\cdot\nabla\partial_s^2)\hat{\mathcal{M}}^{(2k+1)} \\
    &\hat{\mathcal{U}}_{n-1}^{(2k+2)} = i^{1/2}([\fsl{\partial}_s,\fsl{\nabla}]-s\cdot\nabla\partial_s^2)\hat{\mathcal{M}}^{(2k+1)} s\cdot\nabla
\end{align}
while the rest are zero. These satisfy the two conditions $\hat{\mathcal{C}}_j^{(2k+2)}=0$ for any $\hat{\mathcal{M}}^{(2k+1)}$. The second class is seeded by the modifications
\begin{align}
    &\Delta\mathcal{S}_n^{(2k)} = \frac{1}{m^{2k}}\hat{\mathcal{M}}^{(2k)}\Psi_n, && \Delta\mathcal{S}_{n-1}^{(2k)} = \frac{i^{-1/2}}{m^{2k}}\fsl{\partial}_s\hat{\mathcal{M}}^{(2k)}\Psi_n
\end{align}
associated with a gauge invariant operator $\hat{\mathcal{M}}^{(2k)}$ with mass dimension $2k+1$. This introduces $\mathcal{O}(1/m^{2k})$ gauge violations, with associated operators $\hat{\mathcal{O}}_{n-i}^{(2k)}$ and $\hat{\mathcal{U}}_{n-i}^{(2k)}$ equal to
\begin{align}
    &\hat{\mathcal{O}}_{n}^{(2k)} = i^{-1/2}\hat{\mathcal{M}}^{(2k)}s\cdot\nabla \\
    &\hat{\mathcal{O}}_{n-1}^{(2k)} = - i\fsl{\partial}_s\hat{\mathcal{M}}^{(2k)}s\cdot\nabla\\
    &\hat{\mathcal{U}}_{n}^{(2k)} = ([\fsl{\partial}_s,\fsl{\nabla}]-s\cdot\nabla\partial_s^2)\hat{\mathcal{M}}^{(2k)}
\end{align}
while the rest are zero. Again, these satisfy the two conditions $\hat{\mathcal{C}}_j^{(2k)}=0$ for any $\hat{\mathcal{M}}^{(2k)}$. Together, these two classes of interactions can accommodate many types of non-minimal interactions, including multipole and tidal couplings.
\subsubsection{Resummed non-local field strengths}
\label{sec:halfnonlocal}
We now would like to formally resum the modifications $\Delta\mathcal{S}_{n-i}^{(N)}$ and write the interactions in terms of non-local operators.

\sm

Suppose we have field strengths $\mathcal{S}_{n-i}^{(2k)}$ with $\mathcal{O}(1/m^{2k})$ gauge violations that can be cancelled via modifications $\Delta\mathcal{S}_{n-i}^{(N)}$ for all $N > 2k$ in the way described in \autoref{sec:halfRestor}. The gauge invariant field strengths $\mathcal{S}_{n-i} = \mathcal{S}_{n-i}^{(2k)} + \sum_{N = 2k +1}^{\infty}\Delta\mathcal{S}_{n-i}^{(N)}$ may then be written as
\begin{align}
    \mathcal{S}_n & = \mathcal{S}_{n}^{(2k)} +\frac{i}{m^{2k+1}}\hat{\mathcal{O}}_n^{(2k)}\frac{1}{1-\frac{1}{m}s\cdot\nabla\fsl{\partial}_s}\Psi_{n-1}\\
    \mathcal{S}_{n-1} & = \mathcal{S}_{n-1}^{(2k)} - \frac{i^{-1/2}}{m^{2k+1}}\frac{1}{1-\frac{1}{m}\fsl{\nabla}}\Big(\hat{\mathcal{U}}_n^{(2k)}\Psi_n + \hat{\mathcal{U}}_{n-1}^{(2k)}\Psi_{n-1} + \hat{\mathcal{U}}_{n-2}^{(2k)}\Psi_{n-2}\Big) \nonumber\\
    &-\frac{i^{1/2}}{m^{2k+2}}\frac{1}{1-\frac{1}{m}\fsl{\nabla}}\Big(\big(\fsl{\partial}_s(\fsl{\nabla}-m)-s\cdot\nabla\partial_s^2\big)\hat{\mathcal{O}}_{n}^{(2k)}+i\,s\cdot\nabla\hat{\mathcal{O}}_{n-2}^{(2k)}\Big)\frac{1}{1-\frac{1}{m}s\cdot\nabla\fsl{\partial}_s}\Psi_{n-1} \\
    \mathcal{S}_{n-2} & = \mathcal{S}_{n-2}^{(2k)}  +\frac{i}{m^{2k+1}}\hat{\mathcal{O}}_{n-2}^{(2k)}\frac{1}{1-\frac{1}{m}s\cdot\nabla\fsl{\partial}_s}\Psi_{n-1}
\end{align}
One can check that these expressions are gauge invariant and satisfy the covariantized fermionic massive Bianchi identity. To simplify these expressions, we are free to gauge fix, by for instance choosing $\Psi_{n-1}=0$
\begin{align}
    \mathcal{S}_n & = \mathcal{S}_{n}^{(2k)} \\
    \mathcal{S}_{n-1} & = \mathcal{S}_{n-1}^{(2k)} - \frac{i^{-1/2}}{m^{2k+1}}\frac{1}{1-\frac{1}{m}\fsl{\nabla}}\Big(\hat{\mathcal{U}}_n^{(2k)}\Psi_n + \hat{\mathcal{U}}_{n-2}^{(2k)}\Psi_{n-2}\Big) \\
    \mathcal{S}_{n-2} & = \mathcal{S}_{n-2}^{(2k)}
\end{align}

\sm

In this case, it is only for spin $n+1/2 = 1/2$ that the non-local gauge invariant completions vanish. Altogether then, non-local interactions appear to be unavoidable in this construction for massive particles of spin $s\geq 3/2$.
\subsubsection{Spin 3/2}
\label{sec:3/2}
For concreteness, we present here the action for a massive spin 3/2 particle interacting with electromagnetism and gravity resulting from restoring the gauge symmetry after minimal coupling in \autoref{sec:halfRestor}. In this case, the gauge invariant field strengths $\mathcal{S}_1$ and $\mathcal{S}_0$ take the form
\begin{align}
    \mathcal{S}_1 =&\, (\fsl{\nabla} + m - s\cdot\nabla\fsl{\partial}_s)\Psi_1 +i^{1/2}m s\cdot\nabla\Psi_0 + \frac{i^{1/2}}{m}[\fsl{\nabla},s\cdot\nabla]\Psi_0 \\\
    \mathcal{S}_0 =&\, \fsl{\nabla}\Psi_0 + i^{-1/2}m\fsl{\partial}_s\Psi_1 \nonumber\\
    & - \frac{i^{-1/2}}{m}\frac{1}{1-\frac{1}{m}\fsl{\nabla}}([\fsl{\partial}_s,(\fsl{\nabla})^2]-\fsl{\partial}_s[\fsl{\nabla},s\cdot\nabla]\fsl{\partial}_s)\Psi_1 -\frac{1}{m^2}\frac{1}{1-\frac{1}{m}\fsl{\nabla}}\fsl{\partial}_s\fsl{\nabla}[\fsl{\nabla},s\cdot\nabla]\Psi_0
\end{align}
The action is then (\ref{eq:mhalfhyperaction}) at $n=1$
\begin{align}
    S_{3/2} = -\int d^dX \frac{d^dsd^ds'}{(2\pi)^d}e^{is\cdot s'}\Big(&\overline{\Psi}_1\mathcal{S}_1 - \frac{i}{2}\overline{\Psi}_1\overleftarrow{\fsl{\partial}_s}\fsl{\partial}_{s'}\mathcal{S}_1  \nonumber \\
    &+ \frac{i^{-1/2}}{2}\overline{\Psi}_1\overleftarrow{\fsl{\partial}_s}\mathcal{S}_0-\frac{i^{-1/2}}{2}\overline{\Psi}_0\fsl{\partial}_{s'}\mathcal{S}_1 + \frac{1}{2}\overline{\Psi}_0\mathcal{S}_0\Big)
\end{align}
\section{Interpretation of non-local interactions}
\label{sec:fieldredef}
In this section, we discuss possible ways to interpret the non-local interactions. We will provide evidence that there exists a field redefinition which removes the non-local nature of the action. Furthermore, we show that there does not exist a field redefinition which removes the interactions altogether.

\sm 

In carrying out the Noether procedure described in \cite{Berends:1984rq}, we have apparently arrived at non-local interactions for spin $s\geq 3/2$, with non-local length scale $L\sim 1/m$, i.e. the Compton wavelength of the massive particle. Taking this at face value, the infinitely many interactions become relevant at energy scales $E\sim m$, and so there can be no consistent finite truncation of them based on basic effective field theory reasoning. Keeping all such interactions is also of course necessary for gauge invariance. For scattering processes relevant in the classical regime, where the impact parameter $b$ is much larger than the Compton wavelength of any scattered body, any possible non-local effects would be suppressed. Still, this is clearly not an ideal outcome.

\sm

Although the current expressions for the interacting actions are non-local, it is important to consider/emphasize the possibility that the non-locality is in fact fictitious, i.e. can be removed via some complicated field redefinition. Indeed, it is typically assumed that once such non-localities arise in the Noether procedure, interactions can be removed order by order by field redefinition, trivializing the procedure. Imposing locality of the Noether procedure leads to interesting no-go results for interacting massless higher spin fields \cite{Taronna:2017wbx,Roiban:2017iqg}. Before considering whether the non-locality of our interactions can be removed via some field redefinition then, we must first demonstrate that there does not exist a field redefinition which removes the interactions altogether. A useful way to reframe this is to consider whether the interactions have no effect on observables. This can happen if a change in the interaction coupling $g$ in the action can be compensated by a field redefinition, leaving the action unchanged. Such a coupling is called redundant.

\sm

Concretely, if an action $S[\phi]$ depending on some collection of fields $\phi_l(X)$ and couplings $g_i$ satisfies
\begin{align}
\label{eq:redundant}
    \sum_i\frac{\partial S[\phi]}{\partial g_i}\delta g_i = -\epsilon \sum_l\int d^dX\frac{\delta S[\phi]}{\delta \phi_l(X)}F_l(\phi,\partial\phi,\dots)
\end{align}
then this change can be compensated by the infinitesimal field redefinition
\begin{align}
    \phi_l(X) \to \phi_l(X) + \epsilon F_l(\phi,\partial\phi,\dots)
\end{align}

\sm

The right hand side of (\ref{eq:redundant}) is proportional to the equations of motion, and so a coupling $g$ is redundant if the derivative of $S[\phi]$ with respect to $g$ vanishes after the equations of motion are imposed. For the purposes of perturbation theory, it is sufficient to demonstrate that an interaction coupling $g$ is not redundant by showing that the derivative of $S[\phi]$ with respect to $g$ is not zero after setting $g=0$ and imposing the equations of motion at $g=0$. This means that the theory with an infinitesimally small $g$ has different predictions from the free theory with $g=0$. It is through this that we are able to establish that the couplings $e$ and $\kappa$ which appear after making the substitutions $A_{\mu}\to eA_{\mu}$, $g_{\mu\nu}\to \eta_{\mu\nu}-2\kappa h_{\mu\nu}$ in the interactions constructed in this paper are not redundant, and therefore have an effect on observables, at least classically. 

\sm

At $e,\kappa = 0$, the equations of motion are simply those of the free theory, which make $\Phi_{n-1}(X,s),\Phi_{n-2}(X,s),\Phi_{n-3}(X,s) = 0$ and $\Phi_n(X,s)$ a Fierz-Pauli system
\begin{align}
\label{eq:hyperFPs}
    (\partial_X^2-m^2)\Phi_n &= 0 \nonumber\\
    \partial_s\cdot\partial_X\Phi_n &=0 \nonumber\\
    \partial_s^2\Phi_n &=0 
\end{align}
for integer spins, and make $\Psi_{n-1}(X,s),\Psi_{n-2}(X,s)=0$ and $\Psi_n(X,s)$ a Rarita-Schwinger system
\begin{align}
\label{eq:hyperRaritaSchwinger}
    (\fsl{\partial}_X + m)\Psi_n&=0\nonumber\\
    \fsl{\partial}_s\Psi_n&=0
\end{align}
for half integer spins. Taking the derivative of $S_n$ and $S_{n+1/2}$ with respect to $e$, imposing (\ref{eq:hyperFPs}), (\ref{eq:hyperRaritaSchwinger}), and setting $e,\kappa=0$,
\begin{align}
    \frac{\partial S_n}{\partial e}\Big|_{e,\kappa=0} =&\, n!\int d^dX\frac{d^dsd^ds'}{(2\pi)^d}e^{is\cdot s'}\tilde{\Phi}_n\frac{\partial\mathcal{F}_n}{\partial e}\Big|_{e,\kappa=0} = \int d^dX A_{\mu}J^{\mu}_n
\\
    \frac{\partial S_{n+1/2}}{\partial e}\Big|_{e,\kappa=0} =& -n!\int d^dX\frac{d^dsd^ds'}{(2\pi)^d}e^{is\cdot s'}\overline{\Psi}_n\frac{\partial\mathcal{S}_n}{\partial e}\Big|_{e,\kappa=0}= \int d^dX A_{\mu}J^{\mu}_{n+1/2}
\end{align}
where $J^{\mu}_n$ and $J^{\mu}_{n+1/2}$ are the global $U(1)$ currents at $e,\kappa=0$, assuming minimal coupling. These are non-zero, and so turning on minimal electromagnetic interactions in the way presented in this paper cannot be removed by a field redefinition.

\sm

The same is true for gravitational interactions. Taking the derivative of $S_n$ and $S_{n+1/2}$ with respect to $\kappa$, imposing (\ref{eq:hyperFPs}), (\ref{eq:hyperRaritaSchwinger}), and setting $e,\kappa=0$,
\begin{align}
    \frac{\partial S_n}{\partial \kappa}\Big|_{e,\kappa=0} =&\, n!\int d^dX\frac{d^dsd^ds'}{(2\pi)^d}e^{is\cdot s'}\tilde{\Phi}_n\frac{\partial\mathcal{F}_n}{\partial \kappa}\Big|_{e,\kappa=0}= -\int d^dXh_{\mu\nu}T^{\mu\nu}_n
\\
    \frac{\partial S_{n+1/2}}{\partial \kappa}\Big|_{e,\kappa=0} =& -n!\int d^dX\frac{d^dsd^ds'}{(2\pi)^d}e^{is\cdot s'}\overline{\Psi}_n\frac{\partial\mathcal{S}_n}{\partial \kappa}\Big|_{e,\kappa=0}= -\int d^dX h_{\mu\nu}T^{\mu\nu}_{n+1/2}
\end{align}
where $T^{\mu\nu}_n$ and $T^{\mu\nu}_{n+1/2}$ are the energy-momentum tensors at $e,\kappa=0$, assuming minimal coupling. Again, these are non-zero, and so turning on minimal gravitational interactions in the way presented in this paper cannot be removed by a field redefinition.

\sm

Non-minimal interactions constructed in \autoref{sec:nonmin} and \autoref{sec:halfnonmin} have their own couplings $g$, independent of $e$ or $\kappa$, associated with the interaction. The same reasoning shows that as long as the non-minimal operator $\hat{\mathcal{M}}$ does not involve some combination of the equations (\ref{eq:hyperFPs}) or (\ref{eq:hyperRaritaSchwinger}), which is easy to satisfy, the interaction coupling $g$ is not redundant.

\sm

We are now in a position to make a few comments about the possibility of a field redefinition which does not remove the interactions altogether, but removes the non-locality. At least to the order considered, all of the non-local interactions vanish after imposing the equations of motion in the above argument, showing that there \textit{does} exist a field redefinition which removes the non-local interactions to first order in the interaction couplings. This happens because all non-local interactions in the action involve the auxiliary fields, $\Phi_{n-1},\Phi_{n-2},\Phi_{n-3}$ and traces of $\Phi_n$ for integer spins, and $\Psi_{n-1},\Psi_{n-2}$ and $\gamma$ traces of $\Psi_n$ for half integer spins. To show that there exists a field redefinition to all orders in $g_i,e,\kappa$, one must repeat the above argument at finite $g_i,e,\kappa$. 

\sm

Another comment in this context is that the non-locality is rather mild, in the sense that the gauge invariant field strengths $\mathcal{F}_{n-i}$ and $\mathcal{S}_{n-i}$, which are clearly related to the equations of motion, can be expressed as a single non-local operator acting on a local functional of the fields $\Phi_{n-i}$ and $\Psi_{n-i}$, respectively. Indeed, it is straightforward to show for all interactions that
\begin{align}
    \mathcal{F}_{n-1} &= \frac{1}{1-\frac{1}{m^2}s\cdot\nabla(\partial_s\cdot\nabla-\frac{1}{2}s\cdot\nabla\partial_s^2)}\Big(-\frac{i^{-1/2}}{m}(\partial_s\cdot\nabla-s\cdot\nabla\partial_s^2)\mathcal{F}_n+\frac{i}{2m^2}(s\cdot\nabla)^2\mathcal{F}_{n-3}\Big) \\
    \mathcal{F}_{n-2} &= \nonumber \\ &\frac{1}{1-\frac{1}{m^2}(\partial_s\cdot\nabla-\frac{1}{2}s\cdot\nabla\partial_s^2)s\cdot\nabla}\Big(i\big(\partial_s^2-\frac{2}{m^2}(\partial_s\cdot\nabla-\frac{1}{2}s\cdot\nabla\partial_s^2)^2\big)\mathcal{F}_n-\frac{i^{1/2}}{m}s\cdot\nabla\mathcal{F}_{n-3}\Big) \\
    \mathcal{S}_{n-1} &= - \frac{i^{-1/2}}{m}\frac{1}{1-\frac{1}{m}\fsl{\nabla}}\Big(\big(\fsl{\partial_s}(\fsl{\nabla}-m)-s\cdot\nabla\partial_s^2\big)\mathcal{S}_n + i\,s\cdot\nabla\mathcal{S}_{n-2}\Big)
\end{align}
We saw in \autoref{sec:intnonlocal} and \autoref{sec:halfnonlocal} that there are the special gauges $\Phi_{n-1},\Phi_{n-2}=0$ and $\Psi_{n-1}=0$, for which $\mathcal{F}_{n},\mathcal{F}_{n-3}$ and $\mathcal{S}_n, \mathcal{S}_{n-2}$ respectively are local functionals of the remaining fields. For half integer spins, the equations of motion are equivalent to $\mathcal{S}_{n-i}=0$, and so in this gauge the equations of motion are equivalent to local ones. For technical reasons, we are unable to demonstrate that the equations of motion for integer spins are equivalent to $\mathcal{F}_{n-i}=0$ for all $n$, but they are for $n=1$, and based on the preceding evidence we conjecture that this is true for all $n$. Assuming this, the equations of motion in the integer spin case too are equivalent to local ones. 

\sm

As a final comment, a feature of the gauge invariant formulation of massive particles presented in \cite{Lindwasser:2023zwo} is that propagators defined by two different \textit{local} gauge fixing functionals are connected by a \textit{non-local} operator. As a consequence, different local gauge fixing functionals amount to shifting around the non-locality between the vertices and propagators in the Feynman rules. To see how this works explicitly, consider in the absence of interactions the two point correlation functions of the complex gauge invariant linear combinations
\begin{align}
    &A_n = \Phi_n + \frac{i^{1/2}}{m}s\cdot\partial_X\frac{1}{1-\frac{1}{2m^2}(s\cdot\partial_X)^2\partial_s^2}X_{n-1}\\
    &A_{n-3} = \Phi_{n-3} + 3i\partial_s^2\frac{1}{1-\frac{1}{2m^2}(s\cdot\partial_X)^2\partial_s^2}X_{n-1}+\frac{i^{-1/2}}{2m}s\cdot\partial_X\partial_s^2\frac{1}{1-\frac{1}{2m^2}s\cdot\partial_X\partial_s^2s\cdot\partial_X}X_{n-2}\\
    &B_n = \Psi_n+\frac{i^{1/2}}{m}s\cdot\partial_X\frac{1}{1-\frac{1}{m}s\cdot\partial_X\fsl{\partial}_s}\Psi_{n-1}\\
    &B_{n-2} = \Psi_{n-2} - 2i^{-1/2}\fsl{\partial}_s\frac{1}{1-\frac{1}{m}s\cdot\partial_X\fsl{\partial}_s}\Psi_{n-1}-\frac{i^{1/2}}{m}s\cdot\partial_X\partial_s^2\frac{1}{1-\frac{1}{m}s\cdot\partial_X\fsl{\partial}_s}\Psi_{n-1}
\end{align}
with $X_{n-1}=\Phi_{n-1}+\frac{i^{1/2}}{2m}s\cdot\partial_X\Phi_{n-2}$ and $X_{n-2}=\Phi_{n-2}+\frac{i^{-1/2}}{m}s\cdot\partial_X\partial_s^2\Phi_{n-1}$. The two point correlation functions $\langle A_n\tilde{A}_n\rangle$, $\langle A_n\tilde{A}_{n-3}\rangle$, $\langle A_{n-3}\tilde{A}_{n-3}\rangle$, $\langle B_n\overline{B}_n\rangle$, $\langle B_n\overline{B}_{n-2}\rangle$, $\langle B_{n-2}\overline{B}_{n-2}\rangle$ coincide with the propagators in the $\Phi_{n-1},\Phi_{n-2}=0$ and $\Psi_{n-1}=0$ gauge, respectively. Because these expressions are gauge invariant, they may be calculated in any gauge. The propagators in the $\Phi_{n-1},\Phi_{n-2}=0$ and $\Psi_{n-1}=0$ gauge are therefore related to the propagators found in \cite{Lindwasser:2023zwo} via a non-local operation for spins $s\geq 5/2$. As suggested earlier, this is related to the fact that the $\Phi_{n-1},\Phi_{n-2}=0$ and $\Psi_{n-1}=0$ gauge removes a substantial amount of non-local interaction terms in the action. It does not remove all non-local interactions however, and a field redefinition for this purpose is still necessary. Importantly, calculating observables in the gauge in \cite{Lindwasser:2023zwo} must include the non-local interactions which are not present in the $\Phi_{n-1},\Phi_{n-2}=0$ and $\Psi_{n-1}=0$ gauge for the two gauges to be equivalent.

\sm

At least for the low spin examples presented in \autoref{sec:2} and \autoref{sec:3/2}, it is possible to write explicit field redefinitions which localize the actions. This is given by first exhausting all of the gauge symmetry by setting $\Phi_1,\Phi_0 = 0$ and $\Psi_0 = 0$, respectively, and then by making the following field redefinitions
\begin{align}
    &\partial_s^2\Phi_2= (1-\frac{1}{m^2}\nabla^2)(\partial_s^2\Phi_2)', &&\fsl{\partial}_s\Psi_1 = (1-\frac{1}{m}\fsl{\nabla})(\fsl{\partial}_s\Psi_1)'
\end{align}
This is sufficient for localizing the minimally coupled spin 3/2 and 2 actions. Naively, these field redefinitions appear to introduce new propagating modes. Quantum mechanically however, this field redefinition introduces a Jacobian factor in the partition function, which may be represented via ghosts, which by construction account for those modes. Typically in the effective field theory setting, these ghosts can be ignored because they either do not propagate or have a mass larger than the cutoff \cite{Arzt:1993gz}. In this case, the ghosts will have mass $m$, and so cannot be ignored. In some sense, the above field redefinition plays a similar role the gauge symmetry does discussed in the preceding paragraph, in the sense that it shifts around the non-locality between the vertices and propagators in the Feynman rules.

\sm

All of this suggests that the non-local actions constructed in this paper are equivalent to local actions via some field redefinition. More work however needs to be done to establish this. We leave such an analysis for future work.
\section{Comparison with other work}
\label{sec:compare}
There have of course been works in the past, most notably \cite{Zinoviev:2006im,Zinoviev:2008ck,Zinoviev:2009hu,Buchbinder:2012iz}, which have constructed interactions between massive spin $s$ particles and electromagnetism and gravity within the gauge invariant formulation of Zinoviev's \cite{Zinoviev:2001dt}. In those works, they succeeded in cancelling all gauge violations resulting from minimal coupling up to $\mathcal{O}(e)$ or $\mathcal{O}(\kappa)$. In order to cancel the gauge violations to this order, they added both non-minimal interactions \textit{and} non-minimal corrections to the spin $s$ gauge transformations. In contrast, we only had to deform the gauge transformation (\ref{eq:Stueck}) and (\ref{eq:halfStueck}) minimally via the replacement $\partial\to\nabla$, and cancelled all gauge violations by adding non-minimal interactions. 

\sm

The reason for the discrepancy between the two approaches is the following. An additional constraint, not used in this work, when cancelling the gauge violation is to impose a bound on the number of derivatives in the non-minimal interactions at each order in $e$ or $\kappa$. In \cite{Zinoviev:2009hu,Buchbinder:2012iz}, they explicitly place a bound on derivatives at first order in $e$ and find one solution to cancelling the gauge violation with non-minimal interactions and gauge transformations which include up to two derivatives for every spin, and a two parameter space of solutions when including up to three derivatives. As the derivative bound increases, the number of possible non-minimal interactions and gauge corrections increases, and there will in general be a larger parameter space of solutions than the one presented in \cite{Buchbinder:2012iz}. With a finite derivative bound $N$ on the interactions it is necessary to add non-minimal corrections to the gauge transformations, because without them, $\delta S$ will always contain a term with $N+1$ derivatives which does not vanish. As found in the current work, it is only when one does not place a bound on the number of derivatives in the interactions, where formally $N\to\infty$, that one can find a solution which does not require adding non-minimal corrections to the gauge transformations. 

\sm

It is unclear to what extent the actions constructed in this work are related to those constructed in \cite{Zinoviev:2006im,Zinoviev:2008ck,Zinoviev:2009hu,Buchbinder:2012iz}. At face value, the interactions presented in \cite{Buchbinder:2012iz} are bounded in derivatives at first order in $e$, making them starkly different from the interactions in this work. However, it may be that it is untenable to bound derivatives to all orders in $e$ or $\kappa$. Ultimately, the solution space of interactions and gauge transformations which cancel all gauge violations is expected to be large, and the solutions are not necessarily related. A more complete study of the space of solutions, by increasing the bound on derivatives beyond two but still finite, is still needed.

\section{Discussion}
\label{sec:disc}
In this paper, we successfully constructed gauge invariant interactions between massive particles of any spin and electromagnetism and gravity. A non-abelian gauge field extension of these results is straightforward. A conceivable possibility would have been that the massive gauge symmetry predicted unique interactions, perhaps predicting the interactions compatible with tree level unitarity \cite{Ferrara:1992yc,Cucchieri:1994tx}. Instead, we find complete freedom in choosing multipole moments and electromagnetic/gravitational susceptibilities, via the interactions discussed in \autoref{sec:nonmin} and \autoref{sec:halfnonmin}, as well as freedom in choosing all types of interactions linear in $U(1)$ charge 0 integer spin matter fields discussed in \autoref{sec:linearint}. The way in which we have constructed gauge invariant interactions is by no means unique, and their may in fact be for any given value of spin, a simpler formulation than the one arrived in this paper. Indeed, there are explicit examples in the literature which exhibit this non-uniqueness at first order in the Noether procedure \cite{Zinoviev:2006im,Zinoviev:2008ck,Zinoviev:2009hu,Buchbinder:2012iz}.

\sm

The advantage of this construction however is that it works for any spin $s=n,n+1/2$, and any spacetime dimension $d$. When modelling spinning black holes or neutron stars in this way, one must account for classical values of spin, which roughly speaking involves taking the $n\to \infty$ limit while keeping $\hbar n$ finite. Together with the propagators found in \cite{Lindwasser:2023zwo}, it is possible to compute observables as a function of $n$ so that this limit can be taken, while also using dimensional regularization. One challenge to doing this is in taking into account the non-local interactions needed for gauge invariance. 

\sm

It would be interesting to find within this formalism the cubic interactions which reproduce the three-point amplitudes found by Arkani-Hamed--Huang--Huang \cite{Arkani-Hamed:2017jhn} which have ideal high energy behavior, and is known to be related to black hole scattering in the classical large spin limit \cite{Guevara:2017csg,Chung:2018kqs,Guevara:2018wpp,Arkani-Hamed:2019ymq}. Progress on this can be found within other formalisms in for instance \cite{Bern:2020buy, Chiodaroli:2021eug,Skvortsov:2023jbn}. There has also been some interesting work regarding the corresponding classical Compton scattering amplitude \cite{Cangemi:2022bew}, and an uplift to an action principle which reproduces it is still needed. An obstacle is finding consistent quartic interactions which are needed in making the classical limit well-behaved. But at least within this formalism, demanding gauge invariance of these quartic interactions is no longer an obstacle. More generally, it would be interesting to understand, for a given set of cubic interactions, what constraints on the quartic interactions are imposed by demanding a well-behaved classical limit of Compton scattering.

\sm

For spins $s\geq 3/2$, the interactions are non-local, with non-local length scale $L\sim 1/m$. The infinitely many interactions then cannot be ignored at energies $E\sim m$. As discussed in \autoref{sec:fieldredef} however, there is some evidence that there exists a field redefinition which removes the non-locality. An explicit construction of such a field redefinition for $s>2$ is still needed. In the worst case, we expect this construction to still be useful for classical purposes, where the necessary $s\to\infty$ asymptotic limit already precludes a cutoff $\Lambda$ greater than $m$. Here, the priority was to construct a theory which facilitates taking this limit.  

\sm

A low energy effective field theory with energy cutoff $\Lambda$ parametrically larger than the higher spins mass $m$ (a necessary requirement to access the physics of said higher spin particle) will experience causality violations, regardless of the interactions being non-local \cite{Afkhami-Jeddi:2018apj}. String theory offers a causal theory of interacting massive particles with spins $s\geq 2$, but this is only because of finely tuned interactions with other massive particles with a slightly higher mass $m'\gtrsim m$ which pushes the acausal energy scale higher $E\sim m'$. The rough idea is this continues indefinitely, until causality is restored \cite{Camanho:2014apa,Afkhami-Jeddi:2018apj}. In contrast, the theories considered here involve only a single massive particle without self interactions, interacting with electromagnetism and gravity in a generic manner.

\sm

The prospect of constructing self interactions, or interactions between a collection of massive particles of varying types which suppresses causality violations, is a very interesting one. Interactions between massive particles of varying types would require a different gauge principle than the ones presented here. As mentioned above, there is already an explicit example of causal interactions, i.e. the interactions present between massive excitations in string theory. It would be illuminating to reproduce the specific interactions present in string theory from first principles, starting with a free theory and introducing consistent interactions in much the same way it has been done for Yang-Mills theory and general relativity \cite{Berends:1984rq}. 

\sm

Recent progress from bootstrapping scattering amplitudes has shown that there are physically sensible tree level scattering amplitudes involving an infinite collection of massive excitations other than those known to arise from string theory \cite{Geiser:2022exp,Cheung:2022mkw,Cheung:2023adk,Cheung:2023uwn}. A physical realization of these amplitudes is as of yet unknown. The generic nature of these scattering amplitudes offers a clue that the construction of consistent interactions other than the ones present in string theory at the Lagrangian level is possible, providing a physical realization. An analysis of possible interactions would therefore shed light on the uniqueness properties of string theory.

\sm

To explore these possibilities, it is important to gain more control on theories with an infinite collection of massive particles of all representations, not just the totally symmetric spin representations to which we have so far restricted ourselves to. This can for instance be accommodated via a hyperfield $\Phi(X,\{s_i\})$ which is a general function of $N$ auxiliary vector coordinates $s_i^{\mu}$. A formal Taylor expansion in $s_i^{\mu}$ will generate fields $\phi_{\mu_1\cdots\mu_n}^{i_1\cdots i_n}(X)$ which span all Young tableaux if $N > d$. To proceed, one should first construct free actions for massive particles of arbitrary representations which exhibits a gauge symmetry, so that there is a procedure for adding interactions consistent with said gauge symmetry. This will be further expanded on in another work.

\subsection*{Acknowledgements}
The author would like to thank Lucile Cangemi, Henrik Johansson, Paolo Pichini, Massimo Porrati, Trevor Scheopner and E.T. Tomboulis for invaluable discussions during the preparation of this work.  The author would also like to thank Nordita for their hospitality during the workshop ``\href{https://indico.fysik.su.se/event/7916/}{Amplifying Gravity at All Scales}" as part of this work was carried out. Finally, the author thanks Marcus Spradlin for pointing out the possibility of interactions linear in matter fields. This research is supported in part by the Mani L. Bhaumik Institute for Theoretical Physics.

\bibliography{Interactions}

\end{document}